\documentclass[pdflatex,sn-mathphys-num]{sn-jnl}

\usepackage{graphicx}%
\usepackage{multirow}%
\usepackage{amsmath,amssymb,amsfonts}%
\usepackage{amsthm}%
\usepackage{mathrsfs}%
\usepackage[title]{appendix}%
\usepackage{xcolor}%
\usepackage{textcomp}%
\usepackage{manyfoot}%
\usepackage{booktabs}%
\usepackage{algorithmic}
\usepackage{algorithm}%
\usepackage{listings}%
\usepackage{comment}
\usepackage{xspace}
\usepackage{listings}
\usepackage{xcolor}
\usepackage{soul}

\newcommand{\agg}{{\sc Aggregate}\xspace}
\newcommand{\upd}{{\sc Update}\xspace}
\newcommand{\lin}{{\sc Linear}\xspace}

\newcommand{\add}[1]{#1}
\newcommand{\cut}[1]{}

\theoremstyle{thmstyleone}%
\theoremstyle{thmstyletwo}%

\theoremstyle{thmstylethree}%

\begin{document}

\title[FrontGI: Architecture-Aware Parallel GNN Inference]{FrontGI: Architecture-Aware Parallel GNN Inference on Chiplet-Based Multicore Systems}


\author[1]{\fnm{Binglin} \sur{Ji}}\email{binglin.j@wustl.edu}

\author[2]{\fnm{Chenfeng} \sur{Zhao}}\email{chenfeng.zhao@gevernova.com}
\equalcont{Work completed while at Washington~Univ.~in St.~Louis.}

\author*[3]{\fnm{Roger} \sur{Chamberlain}}\email{roger@wustl.edu}

\affil[1]{\orgdiv{Dept.~of Electrical and Systems Engineering}, \orgname{Washington Univ.~in St.~Louis}, \orgaddress{\street{One Brookings Dr.}, \city{St.~Louis}, \state{MO}, \postcode{63130}, \country{USA}}}

\affil[2]{\orgname{GE Vernova Advanced Research}, \orgaddress{\city{Niskayuna}, \state{NY}, \country{USA}}}

\affil[3]{\orgdiv{Dept~of Computer Science and Engineering}, \orgname{Washington Univ.~in St.~Louis}, \orgaddress{\street{One Brookings Dr.}, \city{St.~Louis}, \state{MO}, \postcode{63130}, \country{USA}}}


\abstract{The realm of high-performance computing has expanded to include machine learning tasks of many forms. Graph Neural Networks (GNNs) are becoming increasingly popular, with their applications expanding across diverse domains. As the scale of graph data continues to grow, including larger numbers of nodes, edges, and higher embedding dimensions, standardized libraries such as DGL and PyG have been developed to facilitate GNN computation. These libraries allow the computation to be described a a high level of abstraction, providing high productivity on the part of the developer. However, with the rapid increase in the number of processor cores and the evolution of multicore architectures, these libraries often show poor scalability and fail to execute GNN inference efficiently on the latest multicore systems, particularly those with upwards of a hundred cores. To address this limitation, we introduce FrontGI, a fast and scalable set of techniques that advances the frontier of GNN inference on large-scale graphs (over 100,000 vertices) by leveraging graph-property- and architecture-aware design. FrontGI employs different parallelization strategies optimized for diverse graph structures, maximizing the utilization of multi-level cache hierarchies in multicore systems. We evaluate a number of GNN models with FrontGI on a 128-core AMD EPYC system. Compared to state-of-the-art libraries, FrontGI achieves up to 3.35$\times$, 3.20$\times$, 2.89$\times$ and 3.10$\times$ speedups over DGL for GCN, GraphSAGE, GIN, and SGC models, respectively, and even greater improvements over PyG, across a range of large-scale, high-dimensional graph datasets with diverse structural properties.}




\maketitle

\section{Introduction}
\label{sec:intro}

Recently, large-scale graph data has gained popularity across numerous domains, including social network analysis~\cite{fan2022graph,tabassum2018social}, recommendation systems~\cite{wu2022graph}, bioinformatics~\cite{zhang2021graph}, and electronic circuit design~\cite{dong2023cktgnn}, all of which use high performance computing for their solution. Graph Neural Networks (GNNs) are a specialized type of deep learning model that excels at learning graph data representations. The emergence of GNNs has garnered significant interest from both academic and industrial communities. As the application of GNNs becomes increasingly widespread to computational intensive topics, such as computer vision~\cite{han2022vision} and natural language processing~\cite{wu2021deep, wu2024can}, the dimensionality of GNN embeddings has gradually increased to thousands in recent research. Standard libraries like PyTorch-Geometric (PyG)~\cite{pyg19} and Deep Graph Library (DGL)~\cite{dgl19} are widely used, providing a high-productivity development environment for GNNs. In effect, these libraries allow the application developer to focus on the specifics of the GNN model, removing requirements that the developer port across execution platforms or implement low-level functions. By optimizing sparse operations, in an automated way, these libraries currrently enable users to build and compute graph neural networks with high productivity. One of the clear productivity benefits of libraries like DGL and PyG is that a common source code expression of a GNN application can be efficiently deployed on a wide range of execution platforms. These range from traditional processor cores to a range of computational accelerators.

We propose to expand the scope of execution platforms that libraries like DGL and PyG can exploit to include high core-count multicore systems. Although promising performance results for GNN training and inference are achieved by adopting accelerators such as GPUs~\cite{graphtensor_23,tcgnn_23,pckgnn_24,rtgnn_24}, FPGAs~\cite{geng2020awb,abi2022gengnn,zhao2023gnnhls}, 
and NPUs~\cite{ai2025neutronascend},
multicores remain advantageous and indispensable for GNN execution. 
First, multicores are ubiquitous across all computing platforms, from data centers to personal computers. For instance, since a significant number of cores may be idle during off-peak hours in data centers, leveraging them to perform GNN tasks can substantially reduce the cost~\cite{gong2022graphite}. 
Second, memory capacity poses a significant bottleneck for GNN execution. Large graphs, often reaching hundreds of gigabytes or more, exceed the memory limits of accelerators like GPUs and FPGAs which typically offer only tens of gigabytes of onboard memory. \add{This necessitates frequent data movement between disjoint memories, substantially impacting the resulting performance~\cite{chen2026quantitative}.} In contrast, multicore systems capable of supporting terabytes of memory effectively mitigate this constraint for large-scale GNN tasks~\cite{gong2022graphite}.
Third, multicore platforms can address the poor model scalability observed with discrete GPUs and FPGAs, which suffer from expensive communication bottlenecks~\cite{ma2019neugraph}.
\add{Fourth, while many multicore systems have integrated GPU capabilities, as a general rule, they are substantially less capable than traditional dedicated GPU cards~\cite{ozsoy2020comprehensive}. Finally, libraries such as DGL and PyG offer multicores as target platforms, so improved performance on those platforms can be beneficial to the user community.}

Due to these issues, there is sustained interest in GNN execution on multicore systems, particularly as their computing capability increases.
Several strategies have been proposed to enhance GNN execution on multicores.
GIN~\cite{fu2021automatic} introduces a graph learning method that automatically optimizes and generates computing kernels for GNN inference models based on input graph topology.
Graphite~\cite{gong2022graphite} improves GNN inference through both software and hardware optimizations by fusing operations in each layer, exploiting sparsity from activation functions, and utilizing Direct Memory Access (DMA) engines for data prefetching.
DistGNN~\cite{distgnn21} employs offline graph partitioning methods to reduce communication costs across computing nodes during GNN training.
ARGO~\cite{lin2024argo} uses a Bayesian approach to devise an optimal core-binding strategy, enhancing the scalability of GNN training tasks with standard libraries.
Despite promising results, these approaches typically depend on offline preprocessing, co-design assumptions, or training-specific optimization, and therefore do not directly address a fundamental challenge in GNN inference on modern multicores, especially when offline preprocessing is undesirable or infeasible.
This invites a critical question: 
\emph{how can we effectively accelerate GNN inference by leveraging modern multicore architectures while avoiding offline preprocessing for various graph topologies?} %}

In this work, which extends~\cite{jzc25}, we present FrontGI, a graph-property-aware and architecture-aware set of techniques for GNN inference acceleration on modern multicore processor platforms. The goal is to develop methods that can be incorporated into the existing libraries, maintaining the ability to express the problem at a high level, while enabling efficient implementations to be invoked in an automated way. Here, we significantly broaden the scope of~\cite{jzc25} along three key dimensions. First, motivated by our observation that BLAS parallelization models rely on cache assumptions that do not hold for multicore processors with distributed last-level caches (LLC), FrontGI parallelizes the \upd phase in addition to the \agg phase, which consists of dense matrix operations that dominate the computation in many GNN models. We further investigate how operators can be efficiently fused under distributed LLC architectures. Second, to explore diverse sparse–dense computation patterns, we design and evaluate FrontGI across four representative GNN models: GCN, GraphSAGE, GIN, and SGC. These models exhibit distinct \agg patterns—including additional linear functions and degree-normalized operations—as well as different operator connectivity inside of the \agg operator. Finally, beyond the qualitative analysis, we perform a comprehensive quantitative evaluation using twelve large-scale graph datasets spanning diverse graph sizes, average degrees, and degree distributions, allowing us to systematically characterize how FrontGI selects and adapts parallel execution strategies according to graph structural properties. These strategies are straightforward to incorporate into the standard libraries, maintaining the ability to express the model at a high level of abstraction. Our contributions are as follows:

\begin{itemize}

\item We identify how a distributed last-level cache organization in modern chiplet-based multicores fundamentally influences the performance trade-offs of GNN inference under different graph properties.

\item We show that architectural characteristics impose distinct execution requirements for the \agg phase, and derive two architecture-aware hierarchical execution strategies---hierarchical node-level parallelism and mixed node–tensor parallelism---to mitigate memory access irregularity.

\item We design a simple parallelization strategy for the \upd phase of GNNs that leverages BLAS GEMM kernels to enable faster GNN inference on large-scale graphs with modern multicore architectures, by alleviating the inconsistency between BLAS’s implicit shared-cache access assumptions and the distributed LLC organization of multicore systems.

\item Through extensive evaluation on a wide range of GNN models (GCN, GraphSage, GIN and SGC) and graph datasets with varying properties, we validate our design principles and provide practical guidelines for selecting appropriate parallel execution strategies under different graph structures.

\end{itemize}

GNN inference using the FrontGI techniques achieves a speedup of 3.35$\times$ compared to the maximum performance of the state-of-the-art GNN computation library (DGL). Once incorporated into these libraries, these techniques will continue to improve the performance achievable in a high-productivity development environment.

\section{Motivation and Background}
\label{sec:mbr}

\subsection{Motivation}
\label{subsec:motivation}

It is not uncommon for the performance properties of many applications to change, often significantly, as the core count increases~\cite{gupta1993performance}. Frequently this is due to some property of the architecture or the algorithm that is different for large core counts versus more modest core counts. Systems that allow the expression of applications at a high level of abstraction have been required to automatically adapt to these changes for years. For example, solvers for linear systems are regularly tuned to newly released execution platforms empirically~\cite{whaley2001automated}.

\begin{figure}[htb]
    \centering
    \includegraphics[width=0.67\columnwidth]{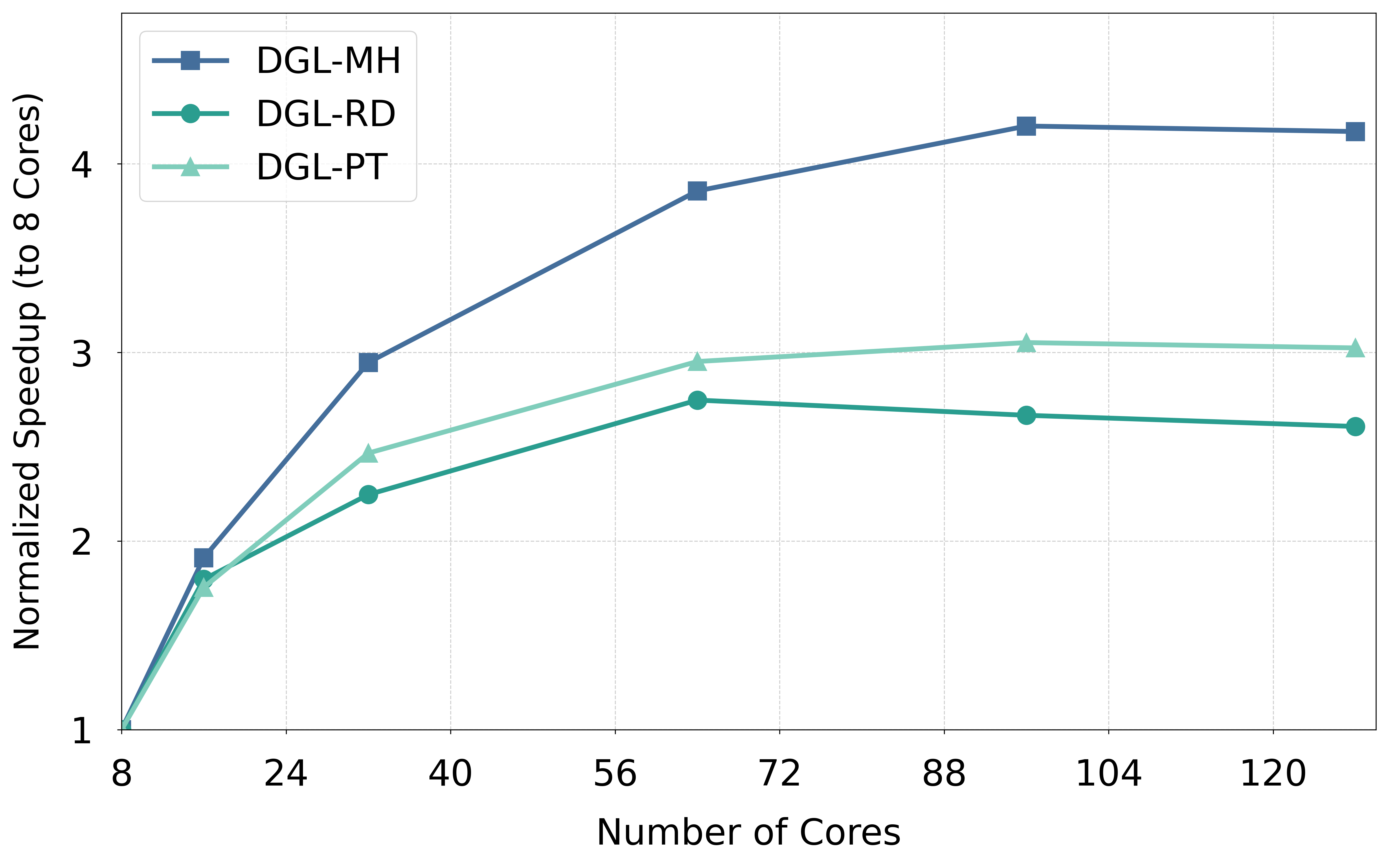}
    \caption{Speedup of DGL 2.4.0 (relative to 8 cores). 
    }
    \label{fig:speedup_dgl_pt}
\end{figure}

Figure~\ref{fig:speedup_dgl_pt} shows the speedup (relative to the 8-core execution) of GCN inference on three graphs with 1024-dimensional embedded features: OGBN-Proteins (PT)~\cite{hu2020open}, Reddit (RD)~\cite{hamilton2017sage} and OGBG-Molhiv (MH)~\cite{hu2020open} with DGL~v2.4.0, executing on a 128-core AMD EPYC system. Up to 32 cores, the parallel efficiency is reasonably good, with each additional core providing significant performance gains. However, the performance gains clearly wane at larger core counts. Specifically, for the OGBG-Molhiv and OGBN-Proteins, performance decreases once the execution exceeds about 90 cores, whereas for the Reddit graph, the decline begins at 64 cores.

\begin{figure}[htbp]
    \hspace*{-0.29cm}
    \centering
    \includegraphics[width=0.69\columnwidth, trim=19 0 18 3, clip]{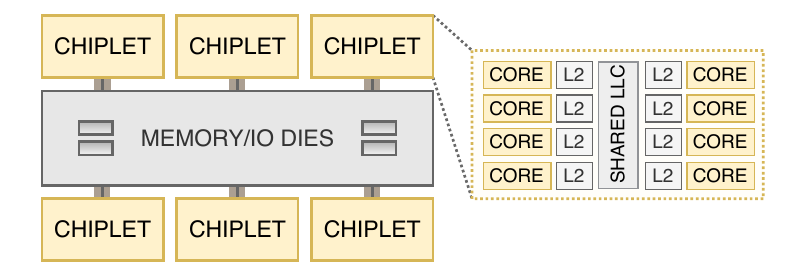}
    \caption{AMD's multicore system using Core Complex Dies (CCDs).}
    \label{fig:ccd}
\end{figure}

One of the potential causes for this rollover in performance is that 128-core systems are architected differently than 16- or 32-core systems. Recently, the designs of high-performance processors have increasingly adopted chiplet-based architectures, driven by the physical and economic limitations of monolithic silicon scaling~\cite{li2020chiplet}. As transistor density growth slows and die sizes approach reticle limits, fabricating large monolithic dies becomes cost-prohibitive and less yield-efficient. To address these challenges, major vendors have adopted chiplet-based designs across various architectures. In multicore processors, AMD employs Core Complex Dies (CCDs) in its EPYC series~\cite{bhargava2024amd}, while Intel uses tile-based designs in processors such as Sapphire Rapids~\cite{nassif2022sapphire}. Take AMD EPYC for instance, the new architecture partitions the multicore processor into multiple chiplets, each integrating a group of physical cores and a dedicated slice of the Last-Level Cache (LLC), interconnected through a high-speed on-package interconnect, as illustrated in Figure~\ref{fig:ccd}. This trend extends beyond traditional processors---many recent accelerator solutions~\cite{smith202411amdgpu} also adopt chiplet architectures to improve scalability.

Data communication within a chiplet is handled through the shared cache, offering low latency, while inter-chiplet communication relies on high-speed interconnects. This architecture significantly improves the scalability and resource utilization of multicore systems through distributed caching and core grouping. However, the backends of standard libraries often leverage OpenMP for data parallelism, typically using graph nodes as the granularity of parallelism. For operators which exhibit highly irregular memory access patterns, common in GNNs, this approach can result in substantial inter-chiplet communication overhead, thereby significantly limiting the scalability of parallel computing. \add{Moreover, for a given hardware platform, the placement of LLC slices across chiplets is entirely determined by the underlying chiplet-based architecture and cannot be controlled directly by software. Consequently, since the cache organization is fixed by the hardware, an important question naturally arises: Can we derive a general parallel execution method that adapts computation to the underlying cache organization? Specifically, given different computation patterns—including varying graph sparsity and sparse–dense data flows—can computation be partitioned in a manner that naturally localizes memory accesses within the hardware-managed LLC of each chiplet while minimizing inter-chiplet communication?}

\add{To address this question and leverage the potential of the new hardware organization, }we explore alternative parallelization techniques for GNNs that attempt to alleviate some of the inter-chiplet communications implied by exclusive use of node-level parallelism. We also measure the impact of these techniques on last-level cache misses. Finally, we articulate how a high-productivity development environment can choose which techniques to deploy.

\subsection{Background}

In this section, we first introduce notation shown in Table~\ref{tbl:notation}, which is used throughout the paper. We then describe different GNN operations and models as well as tensor-program optimizations for neural networks.

\begin{table}[ht]
\caption{Notation.}
\setlength{\tabcolsep}{3pt}
\centering
\renewcommand{\arraystretch}{1.39}
\begin{tabular}{c|l|c|l}
\hline
\textbf{Symbol} & \textbf{Description} 
& \textbf{Symbol} & \textbf{Description} \\ \hline
$\mathcal{G}(\mathcal{V}, \mathcal{E})$ & Graph with nodes $\mathcal{V}$ and edges $\mathcal{E}$ 
& $\mathcal{N}(v)$ & All $v$'s neighbors \\

$\mathrm{V}$ & Node list of $\mathcal{V}$ 
& $\mathrm{E}$ & Edge list of $\mathcal{E}$ \\

$d_v$ & Degree of node $v$ 
& $\mathbf{H}^{(k)}$ & Feature matrix after $k$-th layer \\

$\mathbf{H}_p$ & $p$-th partition of $\mathbf{H}$ 
& $\mathbf{h}_v^{(k)}$ & Node $v$ feature vector at $k$-th layer \\

$\mathbf{A}^{(k)}$ & Aggregated feature matrix at $k$-th layer 
& $\mathbf{A}_p$ & $p$-th partition of $\mathbf{A}$ \\

$\mathbf{a}_v^{(k)}$ & Aggregated feature of node $v$ at $k$-th layer 
& $\mathbf{W}^{(k)}$ & Weight matrix of $k$-th layer \\

$C$ & Number of cores in the system
& $D$ & Number of chiplets in the system\\ \hline
\end{tabular}
\label{tbl:notation}
\end{table}

\subsubsection{Graph Neural Network Operations}
\label{sec:gnn_operations}

Graph Neural Networks (GNNs) compute node feature vectors (embeddings) iteratively on a graph $\mathcal{G}(\mathcal{V}, \mathcal{E})$. 
For a node $v$ at layer $k+1$, its feature vector is derived with two steps based on the embeddings from its neighbors at layer $k$. 

\begin{figure*}[htbp]
    \centering
    \includegraphics[width=0.99\textwidth]{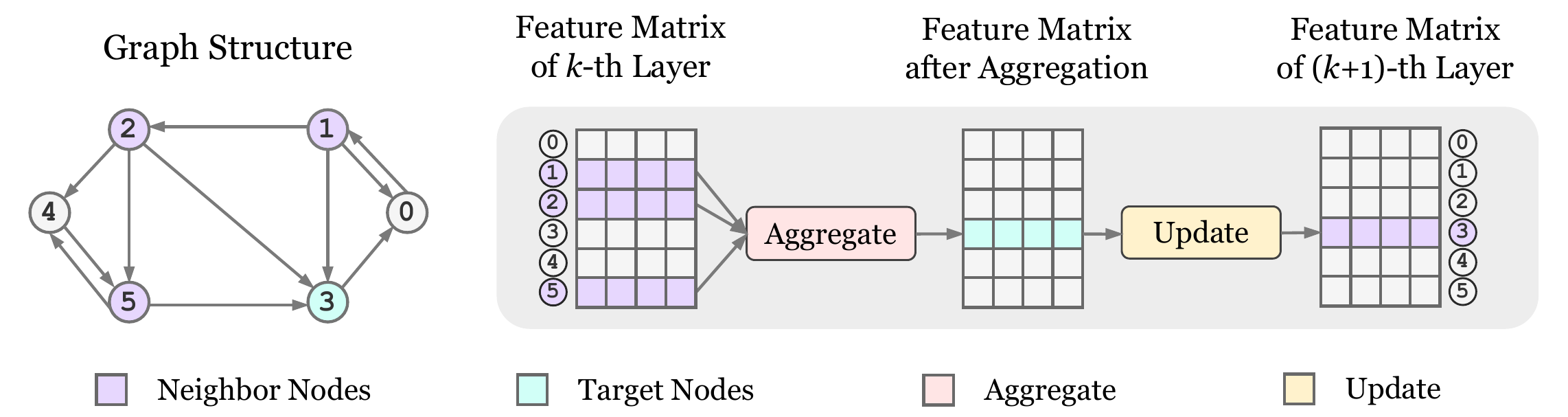}
    \caption{An example graph and corresponding GNN computation  pipeline.}
    \label{fig:gnn_pipeline}
\end{figure*}

The first step is \agg, as shown in Equation~(\ref{eq:aggr}):
\begin{equation}
    \mathbf{a}_v^{(k+1)} = \text{\agg}^{(k+1)}\left(\{\mathbf{h}_u^{(k)} \mid u \in \mathcal{N}(v) \cup \{v\}\}\right)
    \label{eq:aggr}
\end{equation}
%\noindent
where $\mathcal{N}(v)$ represents all neighbors of vertex $v$, $\mathbf{h}_u^{(k)}$ represents node embeddings that need to be aggregated in the $k$th layer. The \agg operator passes messages from $v$'s neighbor nodes, returning aggregated embedding $\mathbf{a}_v^{(k+1)}$.

In the next step, the GNN layer updates $\mathbf{a}_v^{(k+1)}$ with Equation~(\ref{eq:upd}):
\begin{equation}
    \mathbf{h}_v^{(k+1)} = \text{\upd}^{(k+1)}\left(\mathbf{a}_v^{(k+1)}\right)
    \label{eq:upd}
\end{equation}
%\noindent
where the \upd operator is usually a neural network function, a single neural network layer containing a linear and a non-linear (activation) transformation, or a multi-level perceptron. The whole computing pipeline is illustrated in Figure~\ref{fig:gnn_pipeline}.

\subsubsection{Graph Convolutional Network}

The Graph Convolutional Network (GCN)~\cite{gcn} model is one of the most popular GNN models, it sums up all the neighbor node's embeddings in the aggregation step, then updates the aggregated node embeddings with a single neural network layer. Equation~(\ref{eq:gcn}) shows a GCN layer with ReLU activation function applying on a graph node $v$:
\begin{equation}
    \mathbf{h}_v^{(k+1)} = \text{ReLU}\left(\mathbf{W}^{(k)} \sum_{u \in \mathcal{N}(v)} \mathbf{h}_u^{(k)}\right)
    \label{eq:gcn}
\end{equation}
%\noindent
where $\mathbf{W}^{(k)} \in \mathbb{R}^{d_{\text{out}} \times d_{\text{in}}}$  is a learnable weight matrix.

\subsubsection{GraphSAGE}
GraphSAGE~\cite{hamilton2017sage} is a GNN model designed for processing large-scale graphs. It aggregates neighbor embeddings and concatenates the aggregated features with the input node features before applying a learnable transformation and a non-linear activation, enabling the model to learn these two types of information separately and assign different weights to them. Equation~(\ref{eq:gs}) shows a GraphSAGE layer with mean value aggregation and ReLU activation function applying on a graph node $v$:
\begin{equation}
\begin{aligned}
\mathbf{h}_v^{(k+1)} &= \mathrm{ReLU} \left(
\mathbf{W}^{(k)} \, \mathrm{Concat} \left(
\mathbf{h}_v^{(k)},
\frac{1}{|\mathcal{N}(v)|} \sum_{u \in \mathcal{N}(v)} \mathbf{h}_u^{(k)}
\right)
\right) \\
&= \mathrm{ReLU} \left(
\mathbf{W}_a^{(k)} \mathbf{h}_v^{(k)} +
\mathbf{W}_b^{(k)} \frac{1}{|\mathcal{N}(v)|} \sum_{u \in \mathcal{N}(v)} \mathbf{h}_u^{(k)}
\right)
\end{aligned}
\label{eq:gs}
\end{equation}
%
%\noindent
where ${\mathbf{W}}^{(k)} \in \mathbb{R}^{d_{\text{out}} \times 2d_{\text{in}}}$ is a learnable weight matrix. To facilitate parallelization and reduce the memory access footprint, we decompose ${\mathbf{W}}^{(k)}$ into two separate matrices, ${\mathbf{W}}_a^{(k)} \in \mathbb{R}^{d_{\text{out}} \times d_{\text{in}}}$ and ${\mathbf{W}}_b^{(k)} \in \mathbb{R}^{d_{\text{out}} \times d_{\text{in}}}$, which are applied to the input node features and the aggregated features, respectively.

\subsubsection{Graph Isomorphism Network}
The Graph Isomorphism Network (GIN)~\cite{xu2018gin} model is designed to achieve maximum discrimination among graph structures by mimicking the Weisfeiler-Lehman graph isomorphism test~\cite{weisfeiler1968reduction}. GIN employs a sum aggregator, as summation is proven to be more expressive in distinguishing graph structures. After aggregation, GIN applies a learnable transformation using a multi-layer perceptron (MLP) followed by a non-linear activation. This design allows GIN to capture subtle differences in graph topology more effectively. Equation~(\ref{eq:gin}) shows a GIN layer applied on a graph node $v$:
\begin{equation}
\mathbf{h}_v^{(k+1)} = \text{MLP}^{(k)} \Big( (1 + \epsilon^{(k)}) \cdot \mathbf{h}_v^{(k)} + \sum_{u \in \mathcal{N}(v)} \mathbf{h}_u^{(k)} \Big)
\label{eq:gin}
\end{equation}
%
%\noindent
where $\epsilon^{(k)}$ is a learnable or fixed scalar, and $\text{MLP}^{(k)}$ is a multi-layer perceptron.

\subsubsection{Simple Graph Convolution}

The Simple Graph Convolution (SGC)~\cite{wu2019sgc} model removes nonlinearities between layers and collapses intermediate weights into a single linear layer. It performs $K$ rounds of normalized propagation, i.e., aggregating information from the $K$-hop neighborhood, then applies a weight matrix. The normalization parameter for a node $v$ is defined as the inverse square root of its degree $d_v$ after adding a self-loop. Equations~(\ref{eq:sgc_prop}) and~(\ref{eq:sgc}) show the SGC model applied to a node $v$:
\begin{equation}
    \tilde{\mathbf{h}}_v^{(k+1)} = 
    \sum_{u \in \mathcal{N}(v)\cup\{v\}}
    \frac{1}{\sqrt{\tilde{d}_v\,\tilde{d}_u}}\,
    \tilde{\mathbf{h}}_u^{(k)},\qquad k=0,\ldots,K
    \label{eq:sgc_prop}
\end{equation}
\begin{equation}
    \mathbf{h}_v = \mathbf{W}\,\tilde{\mathbf{h}}_v^{(K)}
    \label{eq:sgc}
\end{equation}
where $\tilde{d}_v = d_v + 1$, $\tilde{\mathbf{h}}_u^{(k)}$ is the intermediate aggregated result of node $u$ after $k$-th propagation round,  $\tilde{\mathbf{h}}_u^{(0)} = \mathbf{h}_u^{(0)}$ is the input feature, $\mathbf{h}_v$ is the final output and $\mathbf{W} \in \mathbb{R}^{d_{\text{out}}\times d_{\text{in}}}$ is a learnable weight matrix.

\begin{figure}[htbp]
    \centering
    \includegraphics[width=1.0\columnwidth, clip]{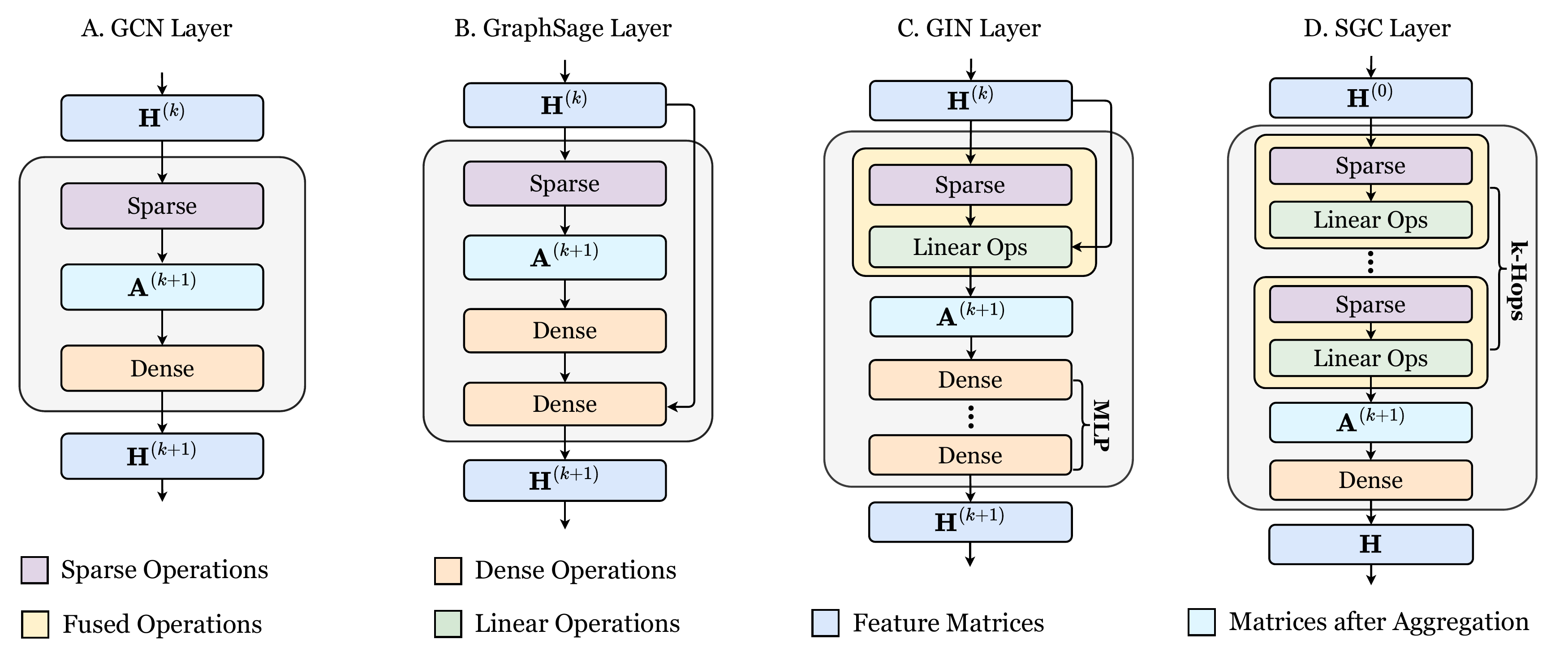}
    \caption{GNN Layer Computation models and dataflows.}
    \label{fig:gnns}
\end{figure}

We select these four models due to their representative dataflow characteristics: GCN and GraphSAGE layers embody the fundamental sparse–dense computation pattern, GIN embodies a sparse-to-multiple-dense computation pattern, while SGC represents multiple sparse operations followed by a dense computation, as illustrated in Figure~\ref{fig:gnns}. As shown in the figure, the dense computations correspond to the update phase, whereas the sparse and fused computations correspond to the aggregation phase.

\subsubsection{Tensor Parallelism for DNNs}

With the increasing scale and dimensionality of datasets for deep learning tasks, tensor parallelism has been proposed to parallelize feature vector computations.
It was first proposed to handle large language model pre-training problems on multiple GPUs by partitioning both the feature and weight matrices' dimensionalities inside of model parallelism~\cite{megatron}, then later was used in high performance AI systems~\cite{tesseract22} and subsequently integrated into standard libraries like PyTorch~\cite{pytorch}. The tensor parallelism method can be illustrated as shown in Algorithm~\ref{alg:tensor_parallelism}.

\begin{algorithm}
\caption{Tensor Parallelism}
\label{alg:tensor_parallelism}
\textbf{Require:} Input tensor $\mathbf{H}$, weight matrix $\mathbf{W}$, number of devices $D$, activation function $\sigma$ 

\textbf{Ensure:} Output tensor $\mathbf{H'}$
\begin{algorithmic}[1]
\STATE Split $\mathbf{H}$ into $[\mathbf{H}_1, \mathbf{H}_2, \dots, \mathbf{H}_D]$ along feature dimension
\STATE Split $\mathbf{W}$ into $[\mathbf{W}_1, \mathbf{W}_2, \dots, \mathbf{W}_D]^T$ along rows
\FOR{$i = 1$ to $D$ \textbf{(in parallel)}}
    \STATE Compute $\mathbf{H}_{temp_{i}} \gets \mathbf{H}_i \cdot \mathbf{W}_i$
\ENDFOR
\STATE Synchronize Partial Results: $\mathbf{H}_{\text{temp}} \leftarrow \sum_{i=1}^{D} \mathbf{H}_{\text{temp}_i}$
\STATE Apply activation function: $\mathbf{H'} \gets \sigma(\mathbf{H}_{temp})$
\end{algorithmic}
\end{algorithm}

When dealing with linear operations, on a system that has $D$ devices, for a batch of data with high dimensional embeddings, we split the input feature matrix $\mathbf{H}$ and weight matrix $\mathbf{W}$ into $D$ parts along feature (column) and row, respectively. Each device processes a partial linear problem. Finally, synchronize and accumulate linear operations, then apply non-linear operations like layer activation.

\section{FrontGI Design}
\label{sec:design}

To push the performance frontier of GNN inference on large-scale graphs using modern multicore processors, we design FrontGI, an efficient and scalable set of computing techniques that synergistically optimizes the computational characteristics of GNNs with the architecture of multicore systems, especially high core count systems built with chiplets. 

Drawing from ideas that work well in FPGA implementations~\cite{zhao2023gnnhls}, we first organize the inference computation in a pipelined manner as Figure~\ref{fig:gnn_pipeline}, where graph data flows from the \agg phase to the \upd phase. While there is a buffer between the two phases to hold intermediate results, this buffer can be relatively small (i.e., it will fit in cache), so that the intermediate results do not have to be written back to main memory. 
In addition, by separating the \agg phase from the \upd phase, instead of executing a large sparse matrix-matrix multiply (SpMM), we are performing a series of vector query and dense vector-matrix multiplies at each node.

To further leverage
the latest multicore architectures, we designed the \agg phase of FrontGI with two different parallel decompositions according to the sparsity of input graphs: mixed tensor/node parallelism for dense graphs and hierarchical node parallelism for sparse graphs. The intermediate results are subsequently stored in a feature matrix buffer, which allows us to more effectively exploit existing matrix libraries during the \upd phase. In both cases, the \upd phase is implemented using the BLAS GEMM library~\cite{dongarra1990set} with our tailored parallelization strategy designed to fully leverage the underlying chiplet-based processor architecture.

\subsection{Graph Processing Systems for GNNs}

General purpose graph processing systems have emerged that handle a wide variety of graph processing problems. Exploiting near-memory processing methods to exploit data locality~\cite{zhao2023supercut}, high efficiency I/O system design to optimize resource usage~\cite{disk12}, and novel graph representation to utilize SIMT architectures~\cite{bfs23,cusha14}, these systems can handle large scale traditional graph processing problems. However, these systems are often not well suited for GNN processing, primarily because most graph processing algorithms' performance is dominated by irregular memory access while GNN processing has both irregular memory access and a high computational load. 
Specifically, \agg involves irregular memory accesses and \upd contains vector-matrix or matrix-matrix multiplication operations that are computationally intensive.

\begin{algorithm}[ht]
\caption{\agg With CSR Format}
\label{alg:aggregate}
\begin{algorithmic}[1]
\REQUIRE Input feature matrix $\mathbf{H}^{(k)}$, feature dimension $dim$, feature buffer $\mathbf{b}_v$, edge list $\mathrm{E}$, edge range $(e_{start}, e_{end})$
\ENSURE Aggregated feature vector $\mathbf{a}_v$ for node $v$
\STATE Initialize $\mathbf{b}_{temp}$ as an array of zeros of size $dim$ 
\FOR{$e = e_{start}$ \TO $e_{end} - 1$}
    \STATE Create a read buffer $\mathbf{b}_{rd}$ as an array of size $dim$
    \STATE $n_{dst} \gets \mathrm{E}[e].\text{src}$
    \FOR{$i = 0$ \TO $dim - 1$}
        \STATE $\mathbf{b}_{rd}[i] \gets \mathbf{H}^{(k)}[n_{dst} \cdot dim + i]$
        \IF{$e = e_{start}$}
            \STATE $\mathbf{b}_{temp}[i] \gets 0$
        \ENDIF
        \STATE $\mathbf{b}_{temp}[i] \gets \mathbf{b}_{temp}[i] + \mathbf{b}_{rd}[i]$
        \IF{$e = e_{end} - 1$}
            \STATE $\mathbf{b}_v[i] \gets \mathbf{b}_{temp}[i]$
        \ENDIF
    \ENDFOR
\ENDFOR
\STATE $\mathbf{a}_v \gets \mathbf{b}_v$

\RETURN $\mathbf{a}_v$
\end{algorithmic}
\end{algorithm}

In this work, by executing \agg and \upd separately in a pipelined manner, we are allowed to tune these two phases individually to address their divergent performance concerns.
Algorithm~\ref{alg:aggregate} illustrates how to execute the \agg operation for a node $v$ with graph data stored in a CSR format (a commonly used format for sparse graphs). The sum operation inherent in GCN's \agg is shown in line~10. When optimizing various GNN models, we substitute this operation with the appropriate alternative.

\subsection{Parallelization of \agg Phase}

\subsubsection{\textbf{Mixed Tensor/Node Parallelism}}

When loading data from memory, a segment of data is loaded into the cache, allowing the processor to exploit data locality. In GNN computing problems, we store node features in a matrix where each row represents a feature vector of a specific node. However, during aggregation, the irregular memory access patterns result in large amounts of data---most of which will not be accessed---being loaded into higher-level caches, causing frequent cache misses due to cache corruption. 
The higher the graph embedding dimension, the better the locality of the data access for feature vectors at an individual core, but correspondingly, the proportion of node features in the cache that will not be accessed also increases.

\begin{figure}[htbp]
    \centering
    \includegraphics[width=0.8\columnwidth]{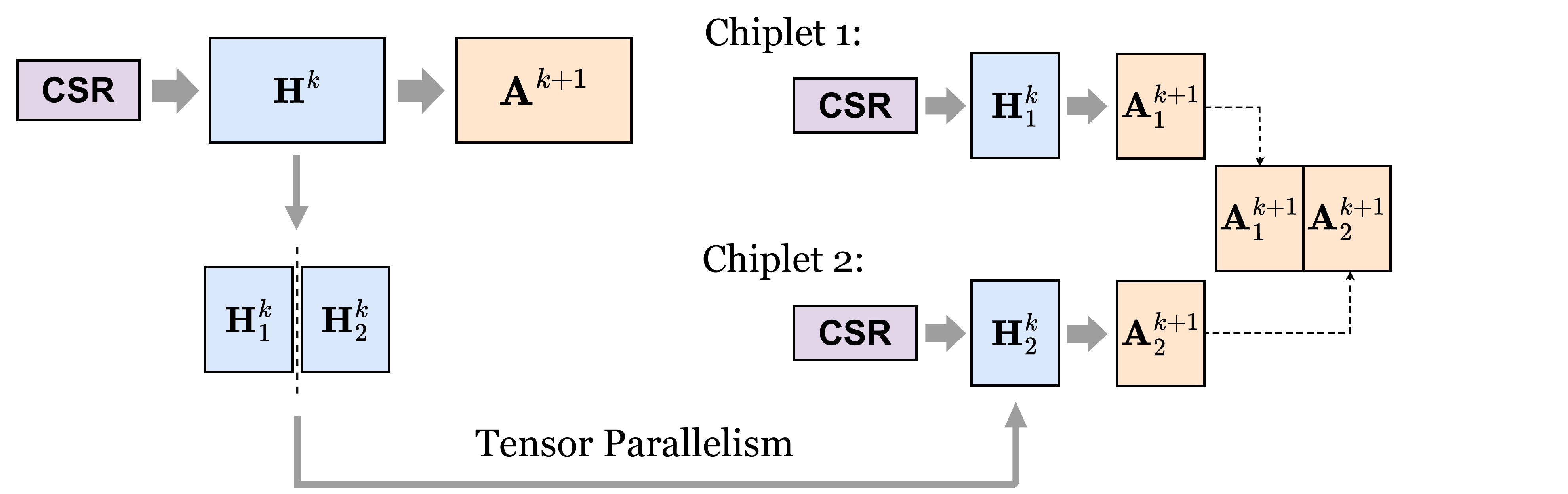}
    \caption{Tensor parallelism in aggregation phase.}
    \label{fig:agg_tensor_para}
\end{figure}

Similar to the tensor parallelism challenge in DNNs, where a single device struggles to handle all features of a batch of data efficiently, in the multicore system scenario, the data segment loading and irregular memory access patterns prevent each chiplet from efficiently handling the full-batch of high-dimensional node features. Therefore, for denser graphs (i.e., graphs with high average node degree), we introduce tensor parallelism during the aggregation phase. Specifically, we apply tensor parallelism to the input feature matrix by treating each chiplet with a shared L3 cache as a ``device.'' Within each chiplet, we exploit node-level parallelism.
In mixed tensor/node parallelism, each chiplet holds its own copy of 
data, recording all nodes' neighbor information and processes in a partial feature matrix along a subset of its dimensions. After all chiplets have processed their corresponding feature subsets, the intermediate results of the aggregation phase are concatenated. Figure~\ref{fig:agg_tensor_para} illustrates applying tensor parallelism to aggregation on a multicore system with two chiplets.  
The left-hand side shows the data decomposition, reading CSR-formatted graph data, and the right-hand side illustrates the mapping to distinct chiplets.

\begin{figure}[ht]
    \centering
    \includegraphics[width=1\columnwidth]{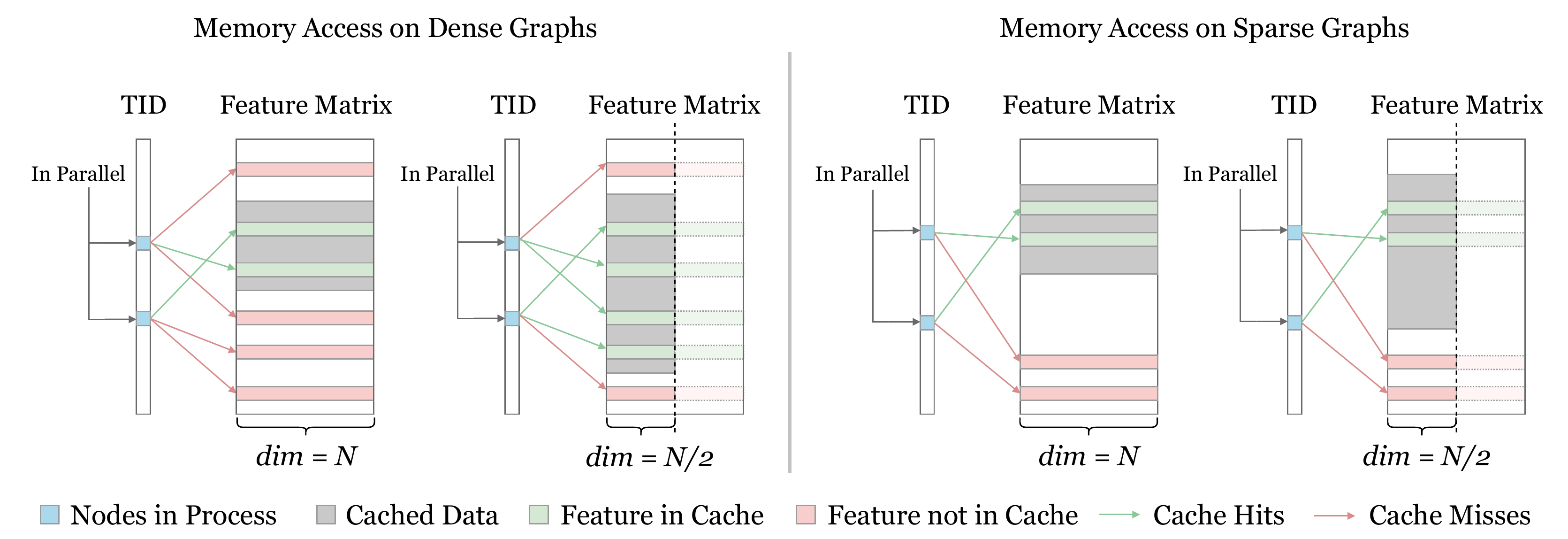}
    \caption{Memory accesses of tensor parallelism on dense (left) and sparse (right) graphs.
    }
    \label{fig:dense_graph}
\end{figure}

In this way, for large core count systems with multiple chiplets, each shared LLC holds a significantly larger portion of node features compared to the pre-tensor-parallelization scenario, thereby reducing LLC load misses during the aggregation phase for denser graphs. As shown in Figure~\ref{fig:dense_graph} (left), when many cores (Core~0 and Core~1) within a chiplet need the feature vectors of the neighbors of the nodes they are currently processing, they first check the cache (here, specifically, the shared LLC within the chiplet). If a cache miss occurs, they will load data from the main memory. By applying tensor parallelism to reduce the dimensionality of the feature vectors, many times more feature vectors than in the pre-tensor-parallelism case can be retained in the cache during the previous data loading process. For dense graphs, this reduces the likelihood of cache misses in LLC. In the figure, the feature size on the left has not been parallelized, so the features occupy more of the cache. On the right, the features are effectively a smaller size, so the cache can hold more of them.

Algorithm~\ref{alg:mixed_aggregation} shows our proposed mixed parallelism for a GNN layer. On a system with $D$ chiplets, for a large input graph with high average node degree and heavy embeddings, we apply tensor parallelism as in Algorithm~\ref{alg:tensor_parallelism} during the \agg phase by splitting the input feature matrix into $D$ parts along the feature dimension. Each chiplet processes a partial aggregation problem with Algorithm~\ref{alg:aggregate}.

\begin{algorithm}[ht]
\caption{Mixed Tensor/Node Parallelism}
\label{alg:mixed_aggregation}
\begin{algorithmic}[1]
\REQUIRE Input feature matrix $\mathbf{H}^{(k)}$, number of chiplets $D$, edge list $\mathrm{E}$, node list $\mathrm{V}$, number of nodes $N$
\ENSURE Output feature matrix $\mathbf{H}^{(k+1)}$
\STATE // Start Aggregate
\STATE $[\mathbf{H}^{(k)}_1, \mathbf{H}^{(k)}_2, \ldots, \mathbf{H}^{(k)}_D] \gets$ split $\mathbf{H}^{(k)}$
\STATE // Tensor Parallelism
\FOR{$d = 1$ \TO $D$ \textbf{in  parallel}}
    \STATE // Compute core IDs for each chiplet
    \STATE $(\mathrm{ID}_{\mathrm{start}}, \mathrm{ID}_{\mathrm{end}}) \gets \text{CoreGroup}(d)$
    \STATE Place all threads in $(\mathrm{ID}_{\mathrm{start}}, \mathrm{ID}_{\mathrm{end}})$

    \STATE // Node Parallelism
    \FOR{$v = 0$ \TO $(N - 1)$ \textbf{in parallel}}
        \STATE Create a buffer $\mathbf{b}_{v}$ with size $dim/D$
        \STATE $(e_{start}, e_{end})  \gets \mathrm{V}(v)$
        \STATE $\mathbf{a}_{v}  \gets \text{\agg}(\mathbf{H}^{(k)}_d, \mathbf{b}_{v}, \mathrm{E}, e_{start}, e_{end})$
        \STATE $\mathbf{A}^{(k+1)}_{d}[v]  \gets \mathbf{a}_{v}$
    \ENDFOR
\ENDFOR
\STATE $\mathbf{A}^{(k+1)} \gets$ concatenate $[\mathbf{A}^{(k+1)}_{1}, \mathbf{A}^{(k+1)}_{2}, \ldots, \mathbf{A}^{(k+1)}_{D}]$
\STATE // Start Update
\STATE $\mathbf{H}^{(k+1)}  \gets \text{\upd}(\mathbf{A}^{(k+1)}, \mathbf{W})$

\RETURN $\mathbf{H}^{(k+1)}$
\end{algorithmic}
\end{algorithm}

\cut{As OpenMP inherently only allows for specifying thread placement policies and does not permit precise control over individual thread placements, we have created a custom thread affinity function.}
\add{Although OpenMP provides policy-based thread affinity mechanisms, our execution strategies require explicit and finer-grained control over the mapping between worker threads and chiplet-local cores. Therefore, we implement a custom thread affinity routine using the Linux \texttt{sched\_setaffinity()} system call. For a two-level nested OpenMP parallel region, each outer worker invokes this routine during the creation of the outer parallel region. The routine constructs a processor affinity mask from the specified $\mathrm{ID}{\mathrm{start}}$ and $\mathrm{ID}{\mathrm{end}}$ values, thereby restricting the worker to a predefined range of cores. The inner OpenMP parallel region is subsequently created within this restricted processor domain, allowing its worker threads to execute only on the assigned cores. Consequently, the operating system enforces the specified CPU affinity and prevents the inner worker threads from migrating outside their assigned processor domain. This mechanism enables us to control thread placement across different stages of GNN execution, facilitating both mixed tensor/node parallelism and hierarchical node parallelism. A simple implementation of the custom thread affinity routine is shown in
Listing~\ref{lst:affinity}.}

\begin{lstlisting}[caption={A
 simple implementation of the thread affinity routine.},label={lst:affinity}]
#include <sched.h>

void set_cpu_affinity(int ID_start, int ID_end) {
    cpu_set_t mask;
    CPU_ZERO(&mask);

    for (int cpu = ID_start; cpu <= ID_end; ++cpu)
        CPU_SET(cpu, &mask);

    sched_setaffinity(0, sizeof(mask), &mask);
}
\end{lstlisting}

Then we parallelize the \agg phase at the node level within this range. Specifically, on a system with $C$ cores, for a feature matrix with dimension $dim$, $\mathbf{H} \in \mathbb{R}^{N \times dim}$, during tensor parallelism, it is divided into $D$ partial matrices $\mathbf{H}_d \in \mathbb{R}^{N \times (dim/D)}$, where $d \in D$. Each chiplet with $\mathrm{ID}_{\mathrm{start}} = (C/D)\times (d - 1)$, and $\mathrm{ID}_{\mathrm{end}} = (C/D)\times d - 1$ processes its own partial matrix with \agg using node parallelism.
With all $\mathbf{H}_d$ processed, we have aggregated feature matrices $\mathbf{A}^{(k+1)}_{d}$, then we concatenate them into a matrix with full feature dimensions $\mathbf{A}^{(k+1)}$, and the \upd function is applied.

\subsubsection{\textbf{Hierarchical Node Parallelism}}

For graphs with low average node degree, tensor parallelism is not well-suited due to two main limitations. First, the overhead of applying tensor parallelism is high. The additional overhead of data partitioning, merging and synchronization in tensor parallelism can take up a large portion of computational cost of the \agg operation, leading to reduced overall performance. Second, the impact of hardware prefetching mechanisms shows significant differences for different graph topologies. For highly regular large embedded graphs, hardware prefetching mechanisms can fully exploit data locality and load node features sequentially into the cache, in which case the benefits of hardware locality are predictable and stable. For highly irregular large embedded graphs, even if the last-level cache can store more node features, due to the irregularity and dispersion of memory access, the node features that need to be aggregated cannot always be loaded into the LLC, and the cache hit/miss problem is difficult to improve significantly. This makes it difficult to guarantee improvement in data locality even when tensor parallelism is used. 

\begin{algorithm}[ht]
\caption{Hierarchical Node Parallelism}
\label{alg:hier_data_aggregation}
\begin{algorithmic}[1]
\REQUIRE Input feature matrix $\mathbf{H}^{(k)}$, number of chiplets $D$, edge list $\mathrm{E}$, node list $\mathrm{V}$, number of nodes $N$
\ENSURE Output feature matrix $\mathbf{H}^{(k+1)}$
\STATE // Start Aggregate, Node Parallelism across chiplets
\FOR{$d = 1$ \TO $D$ \textbf{in  parallel}}
    \STATE // Compute core IDs for each chiplet
    \STATE $ v_{start} = (d - 1) * (N / D) $
    \STATE $ v_{end} = v_{start} + (N / D) $
    \STATE $(\mathrm{ID}_{\mathrm{start}}, \mathrm{ID}_{\mathrm{end}}) \gets \text{CoreGroup}(d)$
    \STATE Place all threads in $(\mathrm{ID}_{\mathrm{start}}, \mathrm{ID}_{\mathrm{end}})$

    \STATE // Node Parallelism across Cores within a chiplet
    \FOR{$v = v_{start}$ \TO $(v_{end} - 1)$ \textbf{in parallel}}
        \STATE Create a buffer $\mathbf{b}_{v}$ with size $dim$
        \STATE $(e_{start}, e_{end})  \gets \mathrm{V}(v)$
        \STATE $\mathbf{a}_{v}  \gets \text{\agg}(\mathbf{H}^{(k)}, \mathbf{b}_{v}, \mathrm{E}, e_{start}, e_{end})$

        \STATE $\mathbf{A}^{(k+1)}[v]  \gets \mathbf{a}_{v}$
    \ENDFOR
\ENDFOR
\STATE // Start Update
\STATE $\mathbf{H}^{(k+1)}  \gets \text{\upd}(\mathbf{A}^{(k+1)}, \mathbf{W})$
\RETURN $\mathbf{H}^{(k+1)}$
\end{algorithmic}
\end{algorithm}

As illustrated in Figure~\ref{fig:dense_graph} (right), when multiple cores process a highly irregular graph with low-degree nodes, applying tensor parallelism indeed results in more feature vectors being loaded in the LLC. However, the number of feature vectors that Core~0 and Core 1 can retrieve from the LLC remains similar to the pre-tensor-parallelism case. Moreover, since tensor parallelism reduces the feature dimensionality, the actual proportion of data that can be hit in the LLC decreases, degrading locality.
\begin{comment}
    
\begin{figure}[ht]
    \centering
    \includegraphics[width=0.8\columnwidth, trim=20pt 15pt 0 20pt, clip]{figures/sparse.pdf}
    \caption{Memory accesses of tensor parallelism on irregular sparse graphs. %\kevin{Updated}
    }
    \label{fig:sparse_graph}
\end{figure}
\end{comment}
Therefore, for sparser graphs, we still apply node parallelism, but optimize the overhead of inter-chiplet communication in a hierarchical way. Based on the analysis of the memory access pattern of graphs with extremely low average node degrees, we adopted a more conservative strategy, \cut{hierarchical node parallelism, as illustrated in Algorithm~4, which is to divide the nodes into $D$ parts according to the node order and assign them to $D$ chiplets in parallel. Each chiplet processes about $N/D$ nodes using Algorithm~2 with node parallelism.}\add{namely hierarchical node parallelism implemented with the same two-level nested parallel execution model. As illustrated in Algorithm~\ref{alg:hier_data_aggregation}, the outer parallel construct partitions the nodes into $D$ parts according to the node order and maps each partition to a chiplet using the custom thread affinity routine. The worker threads within each chiplet then process approximately $N/D$ nodes using Algorithm~\ref{alg:aggregate} with node parallelism in the inner parallel construct.}  

For regular graphs, each node has a limited number of neighbors arranged in a predictable pattern. As a result, when we execute \agg with node parallelism within each chiplet, the feature vectors needed are already loaded or prefetched into the LLCs. For graphs with highly uneven distribution, although the memory access pattern is still irregular, as we fixed certain range of nodes for each chiplet, the variability of executing \agg is reduced compared to single level node parallelism. Threads are scheduled dynamically by OpenMP according to work-load balance regardless of graph structure in single level node parallelism, which brings more uncertainty to feature vectors in each last level cache. When hierarchical parallelism is applied, feature vectors are managed in a more regular way on an inter-chiplet level, ensuring a more deterministic distribution of the feature vectors across chiplets. This more deterministic data distribution may not be optimal, but it provides a baseline for memory access, compared to the less predictable data replacement relying solely on the OpenMP scheduling strategy.

After \agg, the feature vector will be written to the aggregated feature matrix $\mathbf{A}^{(k+1)}$ directly. When all feature vectors are written, \upd is applied.

\subsection{Parallelization of \upd Phase}

\begin{figure}[ht]
    \centering
    \includegraphics[width=0.9\columnwidth]{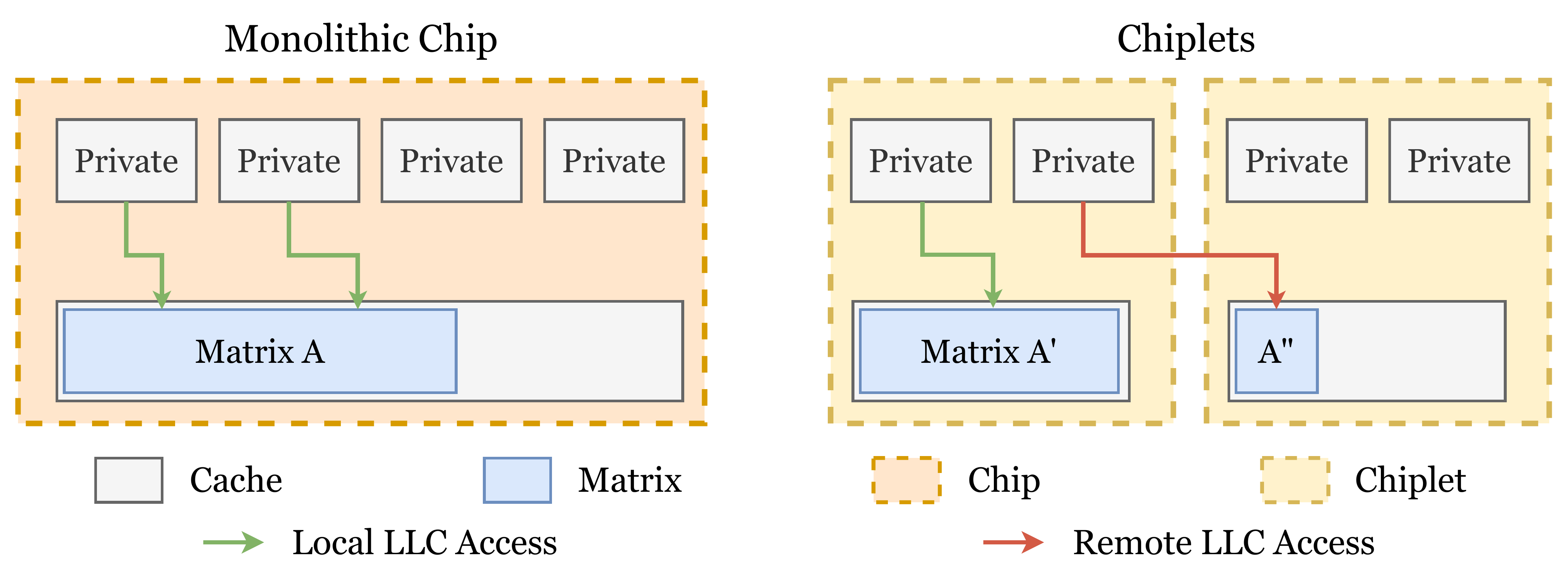}
    \caption{Cache accesses of BLAS on monolithic chip architecture (left) and chiplet-based architecture (right). 
    }
    \label{fig:blas}
\end{figure}

To optimize performance further, we parallelize the \upd phase, i.e., dense neural network layers. Deep learning libraries like PyTorch heavily rely on BLAS as their backend, but BLAS is not specifically optimized for chiplet-based architectures. In the design of classical BLAS~\cite{goto2008anatomy}, the cache hierarchy is modeled as a single logical cache layer, without distinguishing execution units or cache domains. Cache residency is treated as a global modeling state that enables amortization of data reuse, and the model does not assume or specify any form of communication or coherence mechanism between multiple caches. However, on chiplet-based CPUs, reuse across these cache partitions requires explicit cross-LLC communication, which incurs substantial latency and bandwidth overhead; moreover, the access latency from each core’s private cache to the LLC is no longer uniform, contradicting the implicit uniform-LLC-access assumption underlying BLAS kernels. For example, see Figure~\ref{fig:blas}, where matrix $\mathbf{A}$ is loaded into the LLC. On monolithic chips, $\mathbf{A}$ resides in a shared LLC that is directly accessible by the private caches of all cores, allowing private-cache–to–LLC access latency to be modeled uniformly across cores. In contrast, on chiplet-based architectures, $\mathbf{A}$ is partitioned into $\mathbf{A}'$ and $\mathbf{A}''$ and placed in the LLCs of different chiplets; due to the expensive cost of cross-LLC communication, intra-chiplet LLC accesses and inter-chiplet LLC accesses exhibit fundamentally different latencies and cannot be captured by a unified access model. As a result, cache reuse patterns in matrix multiplication that are effective under the classical cache abstraction may lead to significant performance degradation on chiplet-based architectures.

To minimize inter-chiplet communication while still leveraging the highly optimized BLAS kernels at lower cache levels, a simple and effective approach is to execute high-performance BLAS independently within each compute unit that owns private L1–L2 caches. This design avoids modifying BLAS itself, preserves its optimized use of low-level caches, and prevents excessive cache-to-cache traffic caused by the mismatch between the classical BLAS cache abstraction and the physically distributed last-level caches on chiplet-based CPUs. Each unit performs a single \texttt{cblas\_sgemm} call on its assigned block and writes results to the corresponding output region. This design leverages OpenMP to manage parallelism, reducing scheduling overhead.

\begin{algorithm}[htbp]
\caption{Parallel \upd phase across Chiplets}
\label{alg:parallel_sgemm}
\begin{algorithmic}[1]
\STATE \textbf{Input:} Feature matrix $\mathbf{A} \in \mathbb{R}^{N \times d_{\text{in}}}$, weight matrix $\mathbf{W} \in \mathbb{R}^{d_{\text{in}} \times d_{\text{out}}}$, output matrix $\mathbf{H} \in \mathbb{R}^{N \times d_{\text{out}}}$, number of threads $T$
\STATE \textbf{Output:} Updated matrix $\mathbf{H}$
\STATE Divide $N$ rows into $T$ chunks: $chunk\_size \gets \lfloor N / T \rfloor$
\FOR{$t = 0$ \TO $T-1$ \textbf{in parallel}}
    \STATE $r_{\text{start}} \gets t \times chunk\_size$
    \IF{$t = T-1$}
        \STATE $rows \gets N - r_{\text{start}}$
    \ELSE
        \STATE $rows \gets chunk\_size$
    \ENDIF
    \STATE $\mathbf{H}_{r_{\text{start}} : r_{\text{start}}+rows, :} \gets 
       \texttt{BLAS\_GEMM}(\mathbf{A}_{r_{\text{start}} : r_{\text{start}}+rows, :}, \mathbf{W})$
\ENDFOR

\end{algorithmic}
\end{algorithm}

\begin{figure}[ht]
    \centering
    \includegraphics[width=0.7\columnwidth]{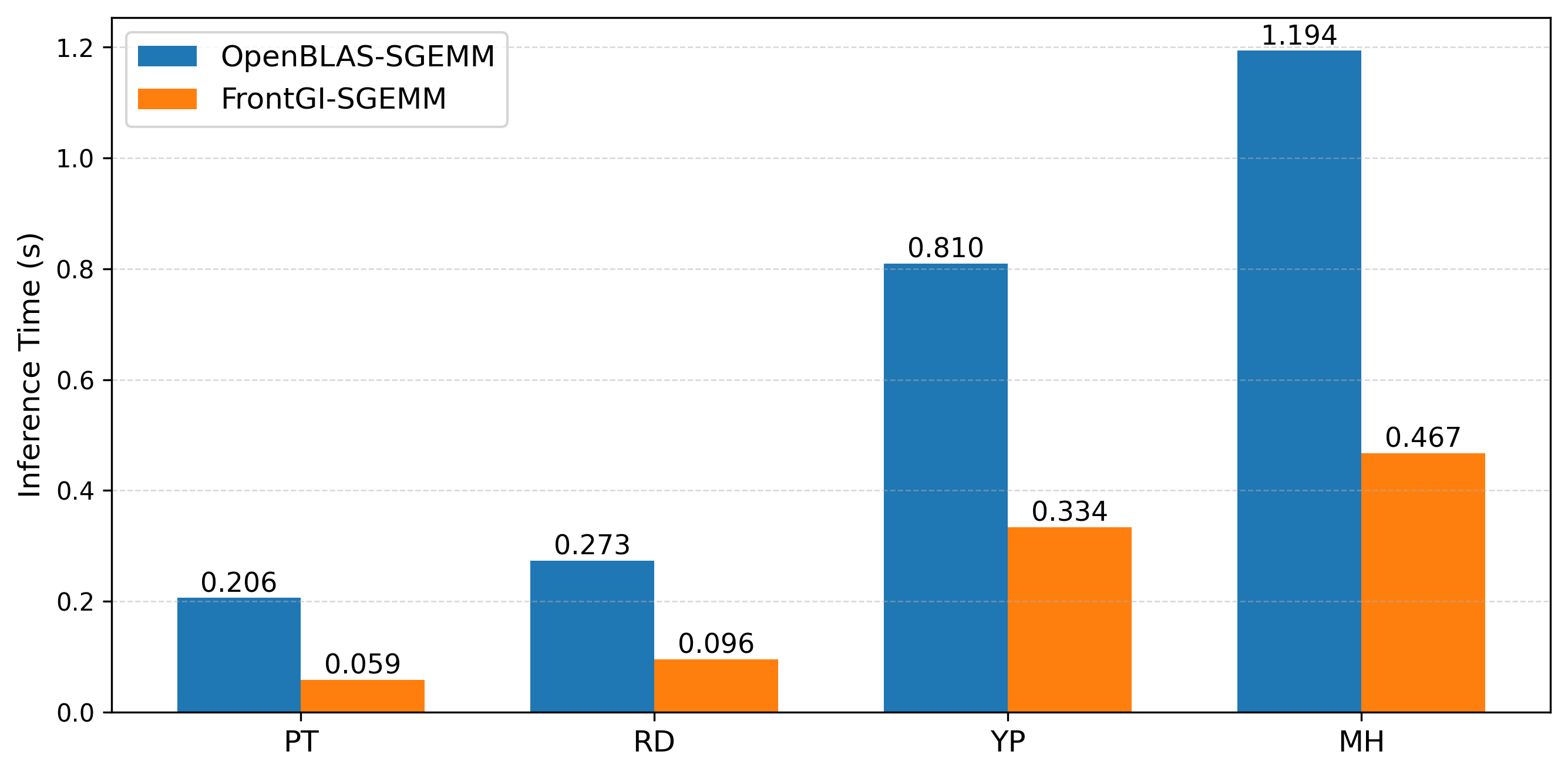}
    \caption{Inference time comparison between OpenBLAS and FrontGI's matrix multiplication in GCN's update phase.}
    \label{fig:blas_comparison}
\end{figure}

The proposed 
strategy is particularly effective when the embedding dimensionality is relatively small compared to the number of nodes, as it improves weight-matrix reuse across multiple row blocks. Moreover, by preserving data locality within each block and maintaining thread affinity, the approach naturally mitigates cross-chiplet data movement—a critical bottleneck in multi-chiplet processor architectures. Figure~\ref{fig:blas_comparison} compares the inference time of matrix multiplication
in the update phase of GCN between OpenBLAS SGEMM and the optimized FrontGI implementation.
All experiments use a feature dimension of 1024 and are evaluated on OGBN-Proteins (132,534 nodes),
Reddit graph (232,965 nodes), Yelp graph (716,847 nodes), and OGBG-Molhiv (1,049,163 nodes). See Table~\ref{tbl:graph_stats} for information on each of the graphs.

\subsection{Fusion of Linear Operators}
When a linear operator exists between the \agg and \upd phases, or is embedded within the \agg phase itself---such as in GIN inference illustrated in Figure~\ref{fig:gnns}, where the aggregated features are combined with the node's initial features through a weighted sum (i.e., an \texttt{AXPY} operation)---we leverage the decomposability and composability of linear operators to fuse this step within the aggregation parallelism. Specifically, during hierarchical node parallelism, each chiplet executes \agg-\lin for approximately $N / D$ nodes. Similarly, under mixed parallelism, after partitioning the input feature dimensions across chiplets, each chiplet performs \agg-\lin on its assigned feature dimensions, followed by a synchronization step to merge partial results.

\subsection{Memory Access Threads Placement}
As discussed in Section~\ref{subsec:motivation}, under a chiplet-based architecture, inter-chiplet communication is relatively costly. Therefore, for memory-intensive operations such as dimension-wise partitioning and concatenation in tensor parallelism, as well as activation functions without complex operations, such as ReLU, we aim to localize these operations within as few chiplets as possible. This approach minimizes inter-chiplet data communication and maximizes data locality. At the same time, we leverage all available threads within these chiplets to exploit parallelism and achieve higher computational efficiency.

\section{Experiments}
\label{sec:experiment}

\subsection{Experimental Methodology}

\subsubsection{Datasets} 
To evaluate our proposed methods, we adopt 12 representative graph datasets from the Open Graph Benchmark~\cite{hu2020open}, Konect Graphs~\cite{kunegis2013konect}, Graph500~\cite{murphy2010graph500} and other public datasets~\cite{hamilton2017sage,zeng2019yelpgraph,dror2012yahoo,lewis2004rtdata} with diverse properties, as shown in Table~\ref{tbl:graph_stats}. 
These graph datasets feature distinct average node degrees ranging from 2.1 to 597.0 and follow either the uniform or power-law~\cite{clauset2009power} degree distribution, indicating diverse irregular memory access patterns which pose unique challenges for graph neural network inference. 
To quantify the degree distribution, we adopt the power-law exponent $\alpha$.
A lower $\alpha$ value indicates a less uniform graph degree distribution, which corresponds to a more irregular memory access pattern, and vice versa.
Typically, highly power-law-like distributions are characterized by  $\alpha < 3$.
Here we note that
$\alpha$ is not applicable to graphs with degree distributions close to standard uniform (e.g., MT and MH shown in Table~\ref{tbl:graph_stats}).
These graphs are converted to CSR format using GNNHLS~\cite{zhao2023gnnhls}.
In addition, we set feature dimensions to be 1024 and adopt 32-bit floating point as the data format, consistent with previous work.

\begin{table*}[ht]
\centering
\caption{Graph Information.}
\label{tbl:graph_stats}
\setlength{\tabcolsep}{4.5pt}
\renewcommand{\arraystretch}{1.2}
\small
\begin{tabular}{cccccccc}
\toprule
\multirow{2}{*}{\textbf{ID}} & \multirow{2}{*}{\textbf{Name}} & \multirow{2}{*}{\textbf{Ref.}} &
\multicolumn{2}{c}{\textbf{Size}} &
\multicolumn{2}{c}{\textbf{Degree}} &
\multirow{2}{*}{\textbf{Power-Law}} \\
\cmidrule(lr){4-5} \cmidrule(lr){6-7}
& & & \textbf{Nodes} & \textbf{Edges} & \textbf{Max.} & \textbf{Avg.} & \textbf{$\alpha$}\\
\midrule
MT & OGBG-Moltox21 & \cite{hu2020open} & 145{,}459 & 302{,}190 & 6 & 2.1 & N/A \\
MH & OGBG-Molhiv & \cite{hu2020open} & 1{,}049{,}163 & 2{,}259{,}376 & 10 & 2.2 & N/A \\
AX & OGBN-Arxiv & \cite{hu2020open} & 169{,}343 & 1{,}166{,}243 & 13{,}155 & 6.9 & 2.56 \\
YP & Yelp Graph & \cite{zeng2019yelpgraph} & 716{,}847 & 13{,}954{,}819 & 4{,}886 & 19.5 & 2.41 \\
KA & Wikipedia (ka) & \cite{kunegis2013konect} & 159{,}756 & 4{,}887{,}982 & 28{,}372 & 30.6 & 2.27 \\
S19 & Graph500-Scale19 & \cite{murphy2010graph500} & 335{,}319 & 15{,}459{,}349 & 41{,}357 & 46.1 & 1.54 \\
PC & Petster-Carnivore & \cite{kunegis2013konect} & 149{,}701 & 10{,}897{,}472 & 80{,}635 & 72.8 & 2.12 \\
HU & Wikipedia (hu) & \cite{kunegis2013konect} & 621{,}512 & 46{,}358{,}909 & 193{,}228 & 74.6 & 2.26 \\
RT & Reuters Graph & \cite{lewis2004rtdata} & 781{,}266 & 121{,}132{,}159 & 345{,}076 & 155.1 & 1.71 \\
YH & Yahoo Music & \cite{dror2012yahoo} & 1{,}000{,}991 & 256{,}804{,}235 & 468{,}366 & 256.8 & 2.34 \\
RD & Reddit Graph & \cite{hamilton2017sage} & 232{,}965 & 114{,}848{,}857 & 21{,}657 & 492.9 & 5.01 \\
PT & OGBN-Proteins & \cite{hu2020open} & 132{,}534 & 79{,}122{,}504 & 7{,}750 & 597.0 & 6.89 \\
\bottomrule
\end{tabular}
\end{table*}

\subsubsection{Evaluation Methods}
To perform the evaluation, we use an AMD EPYC 9754 multicore server featuring 128 cores at 2.25 GHz.
Each core has its own L1d~(32~KiB) cache, L1i~(32~KiB) cache, and L2~(1~MiB) cache. These cores, along with their exclusive caches, are symmetrically distributed across 16 chiplets, with each chiplet sharing a 16 MiB L3 cache.
The system is equipped with 128 GiB of main memory and runs Rocky Linux 9.5.
All code is compiled using GCC 11.5.0 with OpenMP 4.5, and simultaneous multithreading (SMT) is disabled. Therefore, 128 threads are available for model execution.
We compare our methods with two widely-used standard GNN libraries, DGL 2.4.0 and PyG 2.6.1, with PyTorch 2.3.1.
For mixed parallelism during the \agg phase, we split the $\mathbf{H} \in \mathbb{R}^{N \times 1024}$ into 16 partial matrices, each chiplet processes a corresponding submatrix $\mathbf{H}_d \in \mathbb{R}^{N \times 64}$. For hierarchical node parallelism, each chiplet handles $N/16$ nodes. During the \upd phase, we adopt the OpenBLAS library based on OpenMP to perform matrix multiplication, configured to leverage all the 128 cores.

\subsection{Evaluation}

In this section, we present results of our evaluation, including performance, scalability, and LLC hit/miss rates.

\subsubsection{Performance}
We first evaluate performance improvement by showing the overall speedup, defined as the execution time of FrontGI relative to DGL and PyG.
Tables~\ref{tbl:final_gcn_results},~\ref{tbl:final_sage_results}, \ref{tbl:final_gin_results}, and \ref{tbl:final_sgc_results}
show the execution time of DGL and both FrontGI methods on a single GCN, GraphSAGE, GIN and SGC layer, where FrontGI-H refers to Hierarchical Node Parallelism and FrontGI-M refers to Mixed Tensor/Node Parallelism.
For some of the graphs, the table includes execution time for PyG,
where OoM denotes an Out-of-Memory error.

\begin{table}[ht]
\centering
\begin{minipage}{0.49\textwidth}
\centering
\footnotesize
\caption{Execution Time (s) per GCN Layer.}
\label{tbl:final_gcn_results}
\setlength{\tabcolsep}{6pt}
\begin{tabular}{@{}ccccc@{}}
\toprule
\textbf{Name} & \textbf{PyG} & \textbf{DGL} & \textbf{FrontGI-H} & \textbf{FrontGI-M} \\ \midrule
MT & 0.1808 & 0.1550 & \textbf{0.0952} & 0.1492 \\
MH & 1.2745 & 1.0367 & \textbf{0.4923} & 1.0875 \\
AX & 0.2992 & 0.1955 & \textbf{0.1294} & 0.1983 \\
YP & 1.7879 & 0.9806 & \textbf{0.7271} & 0.9139 \\ \addlinespace
KA & 0.5835 & 0.2257 & 0.2081 & \textbf{0.1962} \\
S19 & 1.6908 & 0.5120 & 0.5584 & \textbf{0.4397} \\
PC & 1.6025 & 0.3846 & 0.4736 & \textbf{0.2158} \\
HU & OoM & 1.2663 & 1.1742 & \textbf{0.8178} \\
RT & OoM & 3.3629 & 8.5288 & \textbf{2.0984} \\
YH & OoM & 8.8921 & 24.3678 & \textbf{3.1785} \\
RD & OoM & 1.9101 & 2.7651 & \textbf{0.5859} \\
PT & OoM & 0.8764 & 1.1554 & \textbf{0.2615} \\
\bottomrule
\end{tabular}
\end{minipage}%
\hfill
\begin{minipage}{0.49\textwidth}
\centering
\footnotesize
\caption{Execution Time (s) per GraphSAGE Layer.}
\label{tbl:final_sage_results}
\setlength{\tabcolsep}{6pt}
\begin{tabular}{@{}ccccc@{}}
\toprule
\textbf{Name} & \textbf{PyG} & \textbf{DGL} & \textbf{FrontGI-H} & \textbf{FrontGI-M} \\ \midrule
MT & 0.3320 & 0.2862 & \textbf{0.1527} & 0.2029 \\
MH & 2.2967 & 1.9868 & \textbf{0.9590} & 1.3165 \\
AX & 0.5256 & 0.3427 & \textbf{0.1989} & 0.2685 \\
YP & 2.3794 & 1.8101 & \textbf{1.0098} & 1.2182 \\
KA & 0.7776 & 0.3776 & \textbf{0.2683} & 0.3095 \\ \addlinespace
S19 & 2.0299 & 0.8036 & 0.6773 & \textbf{0.6016} \\
PC & 1.9756 & 0.5162 & 0.5470 & \textbf{0.2438} \\
HU & OoM & 1.8782 & 1.3938 & \textbf{1.0912} \\
RT & OoM & 3.9889 & 8.7681 & \textbf{2.1252} \\
YH & OoM & 9.7020 & 24.4510 & \textbf{3.4640} \\
RD & OoM & 2.0868 & 3.1063 & \textbf{0.6724} \\
PT & OoM & 1.0003 & 1.2602 & \textbf{0.3128} \\
\bottomrule
\end{tabular}
\end{minipage}
\end{table}

\begin{table}[ht]
\centering
\begin{minipage}{0.48\textwidth}
\centering
\footnotesize
\caption{Execution Time (s) per GIN Layer.}
\label{tbl:final_gin_results}
\setlength{\tabcolsep}{6pt}
\begin{tabular}{@{}ccccc@{}}
\toprule
\textbf{Name} & \textbf{PyG} & \textbf{DGL} & \textbf{FrontGI-H} & \textbf{FrontGI-M} \\ \midrule
MH & 2.3629 & 2.0592 & \textbf{0.9508} & 1.3556 \\
AX & 0.5015 & 0.3694 & \textbf{0.2291} & 0.2849 \\
KA & 0.7733 & 0.3920 & \textbf{0.2765} & 0.2797 \\ \addlinespace
PC & 1.8834 & 0.5357 & 0.5246 & \textbf{0.2712} \\
YH & OoM & 9.8108 & 24.1607 & \textbf{3.5115} \\
RD & OoM & 2.0795 & 2.8511 & \textbf{0.7204} \\
\bottomrule
\end{tabular}
\end{minipage}%
\hfill
\begin{minipage}{0.48\textwidth}
\centering
\footnotesize
\caption{Execution Time (s) per SGC Layer.}
\label{tbl:final_sgc_results}
\setlength{\tabcolsep}{6pt}
\begin{tabular}{@{}ccccc@{}}
\toprule
\textbf{Name} & \textbf{PyG} & \textbf{DGL} & \textbf{FrontGI-H} & \textbf{FrontGI-M} \\ \midrule
MH & 2.2995 & 1.5261 & \textbf{0.5886} & 2.1087 \\
AX & 0.8240 & 0.3060 & \textbf{0.1997} & 0.3626 \\
KA & 1.9546 & 0.3765 & \textbf{0.3225} & 0.3501 \\ \addlinespace
PC & 5.7420 & 0.7198 & 0.9028 & \textbf{0.3644} \\
YH & OoM & 17.5066 & 50.0570 & \textbf{6.7910} \\
RD & OoM & 3.5987 & 5.6496 & \textbf{1.1618} \\
\bottomrule
\end{tabular}
\end{minipage}
\end{table}

\begin{figure*}[htb]
    \centering
    \includegraphics[width=0.99\textwidth, trim=10pt 15pt 10 0pt, clip]{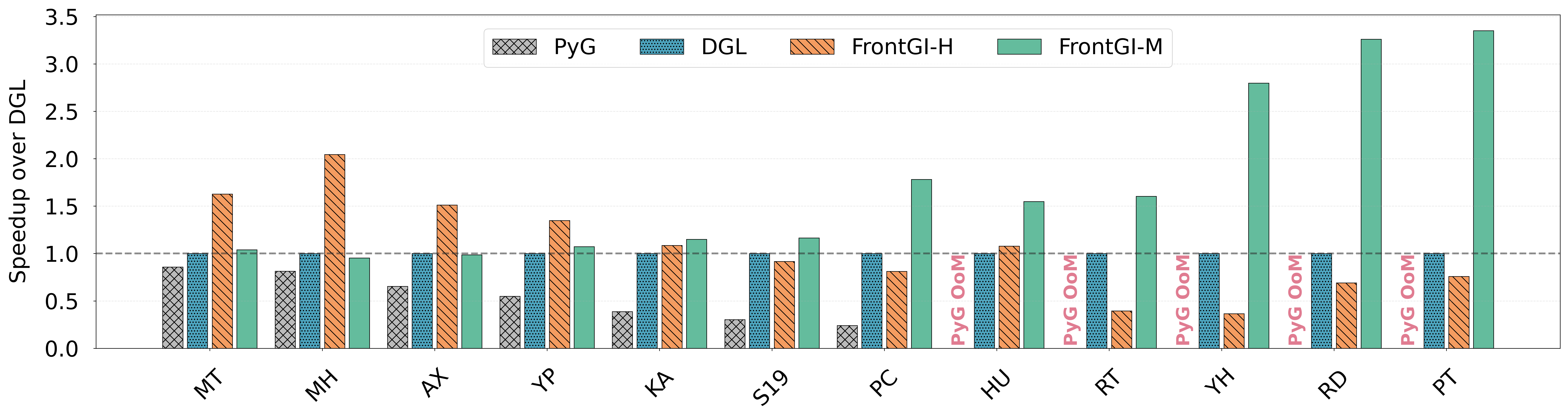}
    \caption{Single GCN layer speedup relative to DGL standard library (Higher is better).}
    \label{fig:speedup_gcn}

    \vspace{2em}
    \centering
    \includegraphics[width=0.99\textwidth, trim=5pt 15pt 5 0pt, clip]{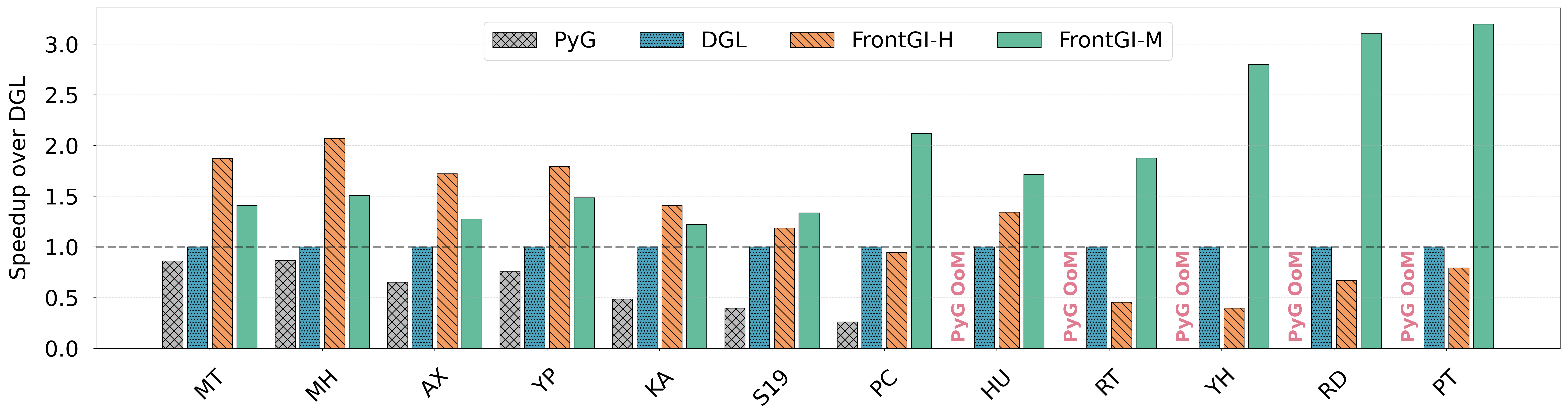}
    \caption{Single GraphSAGE layer speedup relative to DGL standard library.}
    \label{fig:speedup_sage}

    \vspace{2em}

    \begin{minipage}{0.49\textwidth}
        \centering
        \includegraphics[width=\textwidth, trim=0pt 0pt 0 0pt, clip]{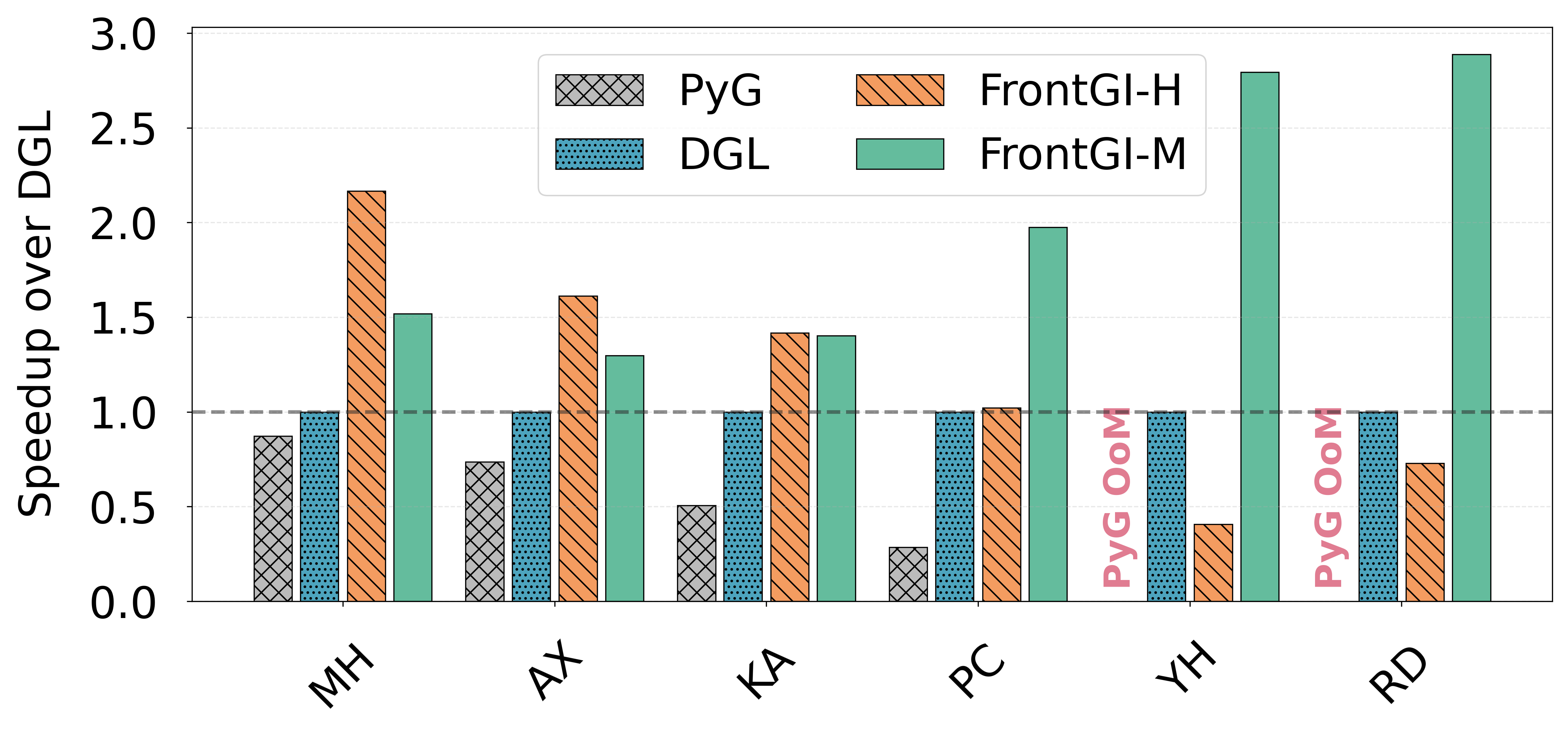}
        \caption{Single GIN layer (with a 2-layer-MLP) speedup relative to DGL.}
        \label{fig:speedup_gin}
    \end{minipage}
    \hfill
    \begin{minipage}{0.49\textwidth}
        \centering
        \includegraphics[width=\textwidth, trim=0pt 0pt 0 0pt, clip]{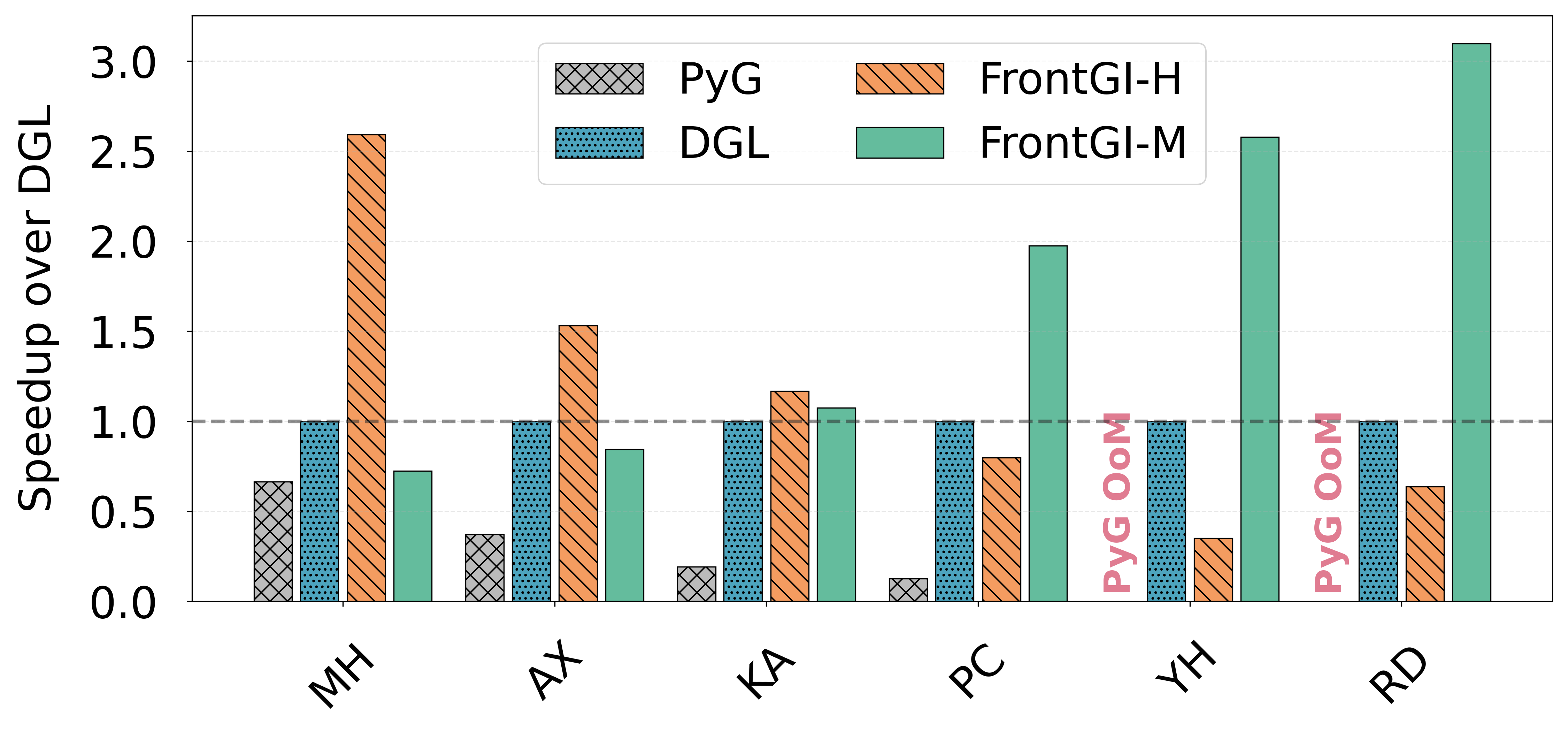}
        \caption{SGC model (with $K$ = 2) speedup relative to DGL.}
        \label{fig:speedup_sgc}
    \end{minipage}
\end{figure*}

Figures~\ref{fig:speedup_gcn}, \ref{fig:speedup_sage}, \ref{fig:speedup_gin}, and~\ref{fig:speedup_sgc} respectively show the speedup 
for GCN, GraphSAGE, GIN, and SGC achieved by PyG standard library and our proposed FrontGI methods, including mixed tensor/node parallelism and hierarchical node parallelism, relative to DGL.

In comparison with DGL, a common trend is observed. For graphs with lower average degrees, FrontGI's hierarchical node parallelism outperforms
DGL, achieving speedup from $1.35\times$ to $1.82\times$ for a GCN layer, $1.72\times$ to $2.07\times$ for a GraphSAGE layer, $1.42\times$ to $2.17\times$ for a GIN layer with a 2-layer MLP, and $1.17\times$ to $2.59\times$ for a SGC model with 2 rounds within \agg.
In contrast, for graphs with high average node degree, the mixed tensor/node parallelism method delivers superior performance, reaching approximately $3.35\times$ the inference speed of DGL for GCN, $3.20\times$ for GraphSAGE, $2.89\times$ for GIN and $3.10\times$ for SGC. 

When compared to PyG, both FrontGI strategies outperform its implementation (on PC graph, FrontGI achieved up to $7.42\times$, $8.10\times$, $6.94\times$ and $15.76\times$ with different models), since PyG’s performance is consistently far below that of DGL and it frequently runs into out-of-memory errors on large graphs with high average degree.

\begin{table}[ht]
\centering
\caption{Profiling a Single GCN Layer on 128 Cores.}
\label{tbl:gcn_profile}
\begin{tabular}{@{}c|cc|ccc@{}}
\toprule
\textbf{Name} & \multicolumn{2}{c|}{\textbf{Hierarchical Node}} & \multicolumn{3}{c}{\textbf{Mixed Tensor/Node}} \\ 
\cmidrule(lr){2-3} \cmidrule(lr){4-6}
 & \agg & \upd & \agg & \upd & \textbf{Overhead} \\ 
\midrule
MT  & \textbf{0.0185} & 0.0749 & 0.0419 & 0.0628 & 0.0422 \\
MH  & \textbf{0.0966} & 0.3940 & 0.3504 & 0.4672 & 0.2578 \\
AX  & \textbf{0.0491} & 0.0800 & 0.0753 & 0.0708 & 0.0493 \\ \addlinespace
YP  & 0.4278 & 0.2896 & \textbf{0.4201} & 0.3338 & 0.1521 \\
KA  & 0.1166 & 0.0879 & \textbf{0.0782} & 0.0677 & 0.0480 \\
S19 & 0.3915 & 0.1638 & \textbf{0.1935} & 0.1650 & 0.0771 \\
PC  & 0.3850 & 0.0817 & \textbf{0.0772} & 0.0934 & 0.0432 \\
HU  & 0.8993 & 0.2933 & \textbf{0.3895} & 0.2896 & 0.1316 \\
RT  & 8.2007 & 0.3197 & \textbf{1.4582} & 0.3632 & 0.2674 \\
YH  & 23.9598 & 0.4010 & \textbf{2.4293} & 0.4496 & 0.2864 \\
RD  & 2.6429 & 0.1194 & \textbf{0.4239} & 0.0956 & 0.0651 \\
PT  & 1.0845 & 0.0691 & \textbf{0.1628} & 0.0586 & 0.0382 \\
\bottomrule
\end{tabular}
\end{table}

Since the irregularity of GNN models is mainly reflected in the \agg phase, we next focus our evaluation on this phase for both proposed methods across all the graph datasets.
Table~\ref{tbl:gcn_profile} shows the execution time of the \agg phase and the \upd phase for mixed parallelism and hierarchical node parallelism.

\begin{figure}[htb]
  \centering
  \includegraphics[width=0.7\columnwidth, trim=0pt 0pt 0 0, clip]{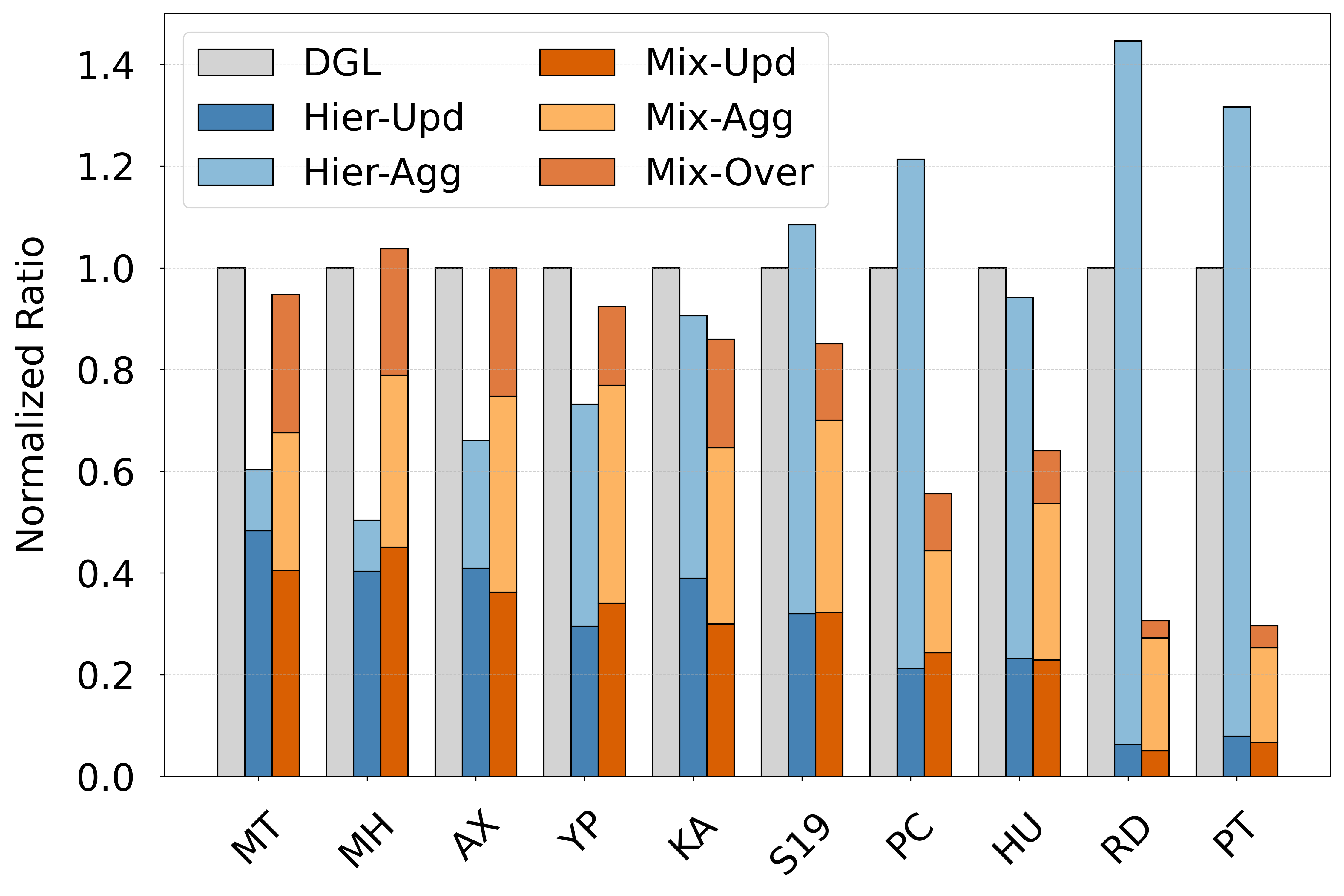}
    \caption{Normalized GCN layer profiling (for each dataset, DGL = 1).}
    \label{fig:prof_gcn}
\end{figure}

In particular, the last column $\textbf{Overhead}$ means the overhead of implementing tensor parallelism within mixed parallelism, i.e., partitioning $\mathbf{H}_{d}^{(k)}$ and merging $\mathbf{A}_{d}^{(k+1)}$. Figure~\ref{fig:prof_gcn} shows the normalized inference time and the comparison between different techniques. The comparison shown in the table and the figure confirms our assumption regarding the impact of processor architecture features on irregular memory access.

When mixed parallelism is applied, the \agg\ time for graphs with an average degree larger than 19 (as in YP) begins to decrease. When the average degree reaches approximately 46 (as in S19), the \agg\ time is reduced significantly enough to offset the additional overhead introduced by tensor parallelism. Additionally, we calculate the overhead-to-gain ratio $R_{\text{OTG}}$ when mixed parallelism outperforms hierarchical node parallelism, which quantifies the proportion of overhead of mixed parallelism relative to the performance improvement as follows: 

\begin{equation}
R_{\text{OTG}} = \frac{t_{\text{overhead}}}{t_{\text{gain}}} = \frac{t_{\text{overhead}}}{t_{\text{hierarch}} - (t_{\text{mixed}} - t_{\text{overhead}})}
\end{equation}
where $t_{\text{overhead}}$ is \textbf{Mixed Overhead} in Table~\ref{tbl:gcn_profile}, $t_{\text{gain}}$ stands for the benefit obtained from mixed parallelism compared to the best performance of hierarchical parallelism. We calculate $R_\text{OTG}$ only when $t_{gain}$ is positive. $t_{\text{hierarch}}$ is the minimum execution time of \agg with hierarchical node parallelism across different core number settings, $t_{\text{mixed}}$ is the execution time of \agg with mixed tensor/node parallelism. Comparing to tensor parallelism's benefit, its overhead is acceptable ($R_{\text{OTG}}=3.96\%$ for RT, $1.33\%$ for YH, $4.14\%$ for PT and $2.93\%$ for RD). 
For sparser graphs, the time taken by tensor parallelism's overhead is comparable to the \agg inference time of hierarchical node parallelism in the 128-core setting, and it can take up to 2.46$\times$ as long as hierarchical node parallelism \agg in the optimal 32-core setting on graphs with regular memory access.

\subsubsection{Impact of Graph Properties}

We observe that the choice of the best-performing parallelization strategy is strongly correlated with the average node degree of the graph. For graphs such as MT, MH, AX, and YP, all of which have an average degree less than 20, hierarchical node parallelism clearly performs the best. For KA and S19, whose average degrees range from about 20 up to around 50, hierarchical node parallelism and mixed tensor/node parallelism perform similarly. In contrast, for graphs like PC, HU, RT, YH, RD, and PT, which have average degrees greater than 50, mixed tensor/node parallelism significantly outperforms hierarchical node parallelism. This is counter to conventional wisdom, in which one assumes that the distribution of the graph degree would have a large performance impact. Here, MT and MH are fairly uniform in distribution, while S19 and RT are quite non-uniform (low $\alpha$ power-law exponent). Yet, this factor was not what distinguished the best performing parallelization technique.

\subsubsection{Scalability}

\begin{figure}[htb]
    \centering
    \includegraphics[width=0.7\columnwidth]{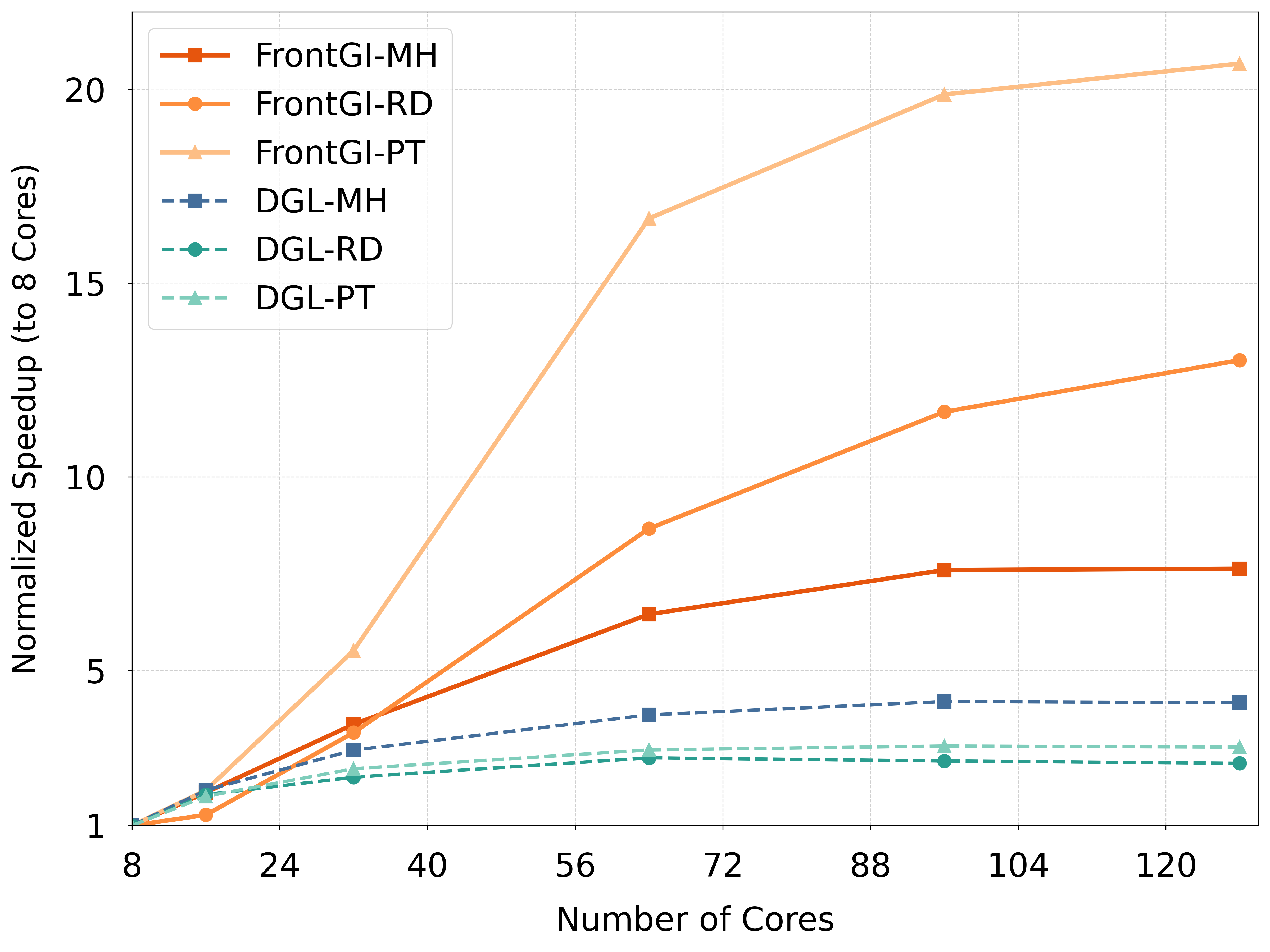}
    \caption{Comparison of FrontGI and DGL scalability on PT, RD and MH graphs (relative to 8 cores).
    }
    \label{fig:scalable_FGI}
\end{figure}

Figure~\ref{fig:scalable_FGI} shows the scalability of FrontGI (hierarchical node parallelism for MH and mixed tensor/node parallelism for PT and RD) and repeats the scalability of DGI from Figure~\ref{fig:speedup_dgl_pt}. The inference speed of 8 cores (for both FrontGI and DGL on each graph) is normalized to 1. In this experiment, we scale the system by progressively adding chiplets rather than merely increasing core counts.
This setup directly follows our chiplets-aware system design, ensuring that the scalability evaluation reflects realistic architectural expansion.
Comparing to Figure~\ref{fig:speedup_dgl_pt}, the performance of FrontGI significantly surpasses DGL with increasing core counts. When scaling up to 32 cores, the speedup relative to 8 cores for DGL begins to plateau, while FrontGI continues to exhibit rapid performance growth until the growth rate starts to decrease when scaling above 64 cores. When the number of cores exceeds 96, DGL's performance begins to decline, while FrontGI maintains and keeps improving its performance up to the full utilization of 128 cores.  
FrontGI's scalability is without question an improvement on that of the standard library.

\subsubsection{LLC Hit/Miss}

In modern multicore systems, LLC cache accesses have a significant impact on the communications between chiplets. We believe that for multicore architectures with multiple chiplets designs, programming paradigms like OpenMP, which implement fork-join parallelism through simple pragmas, are likely to cause a large number of LLC misses due to the lack of thread-level control at the chiplet level. Moreover, LLC misses can result in the most significant cache miss latency and affect cache miss events in other levels. 

In our experiments 
we measure the LLC hit and miss events for each of three parallel methods, all using the FrontGI code base: traditional node parallelism, mixed tensor/node parallelism, and hierarchical node parallelism. 
To facilitate analysis of these data for both high and low degree graphs, we will define a cache ``event'' change ratio, where what constitutes an event is different for distinct comparisons. For all definitions of an event, the change ratio tells us how much FrontGI is different than traditional node parallelism for that type of event.
We define the cache event change ratio $R_{\text{Event}}$ as follows:
\begin{equation}
R_{\text{Event}} = \frac{\left| \text{Event}_{\text{FrontGI}} - \text{Event}_{\text{Node}}\right|}{\text{Event}_{\text{Node}}}
\end{equation}
where $\text{Event}$ represents one of: (1)~cache hits, (2)~cache misses, or (3)~cache accesses (the sum of hits and misses).

Table~\ref{tbl:cache_miss_mixed} shows cache hit and miss events on high degree graphs before and after applying tensor parallelism (i.e., for pipelined node-level parallelism and mixed tensor/node parallelism). 
\begin{table}[ht]
\centering
\caption{LLC Hit/Miss counts For High Degree Graphs.}
\label{tbl:cache_miss_mixed}
\begin{tabular}{@{}cccccc@{}}
\toprule
 & \multicolumn{2}{c}{\textbf{L3 hit}} & \multicolumn{2}{c}{\textbf{L3 miss}} \\ \cmidrule(l){2-5} 
 & \textbf{Node} & \textbf{FrontGI-M} & \textbf{Node} & \textbf{FrontGI-M} \\ \midrule

S19 & 622\,M & 906\,M & 3\,575\,M & 1\,057\,M \\
PC & 918\,M & 434\,M & 1\,543\,M & 503\,M \\
HU & 2\,430\,M & 1\,013\,M & 6,336\,M & 2,302\,M \\
RT & 4\,746\,M & 3\,337\,M & 12\,494\,M & 6\,267\,M \\
YH & 3\,149\,M & 11\,460\,M & 26\,812\,M & 12\,578\,M \\
PT & 3\,003\,M & 4\,280\,M & 5\,287\,M & 659\,M \\
RD & 2\,053\,M & 4\,333\,M & 9\,874\,M & 2\,417\,M \\ \bottomrule
\end{tabular}
\end{table}
Looking first at cache misses ($\text{Event}=\text{Miss}$), we observe a significant decrease ($R_{\text{Miss}} = 70.43\%$ on S19, 67.40\% on PC, 63.67\% on HU, 49.83\% on RT, 
53.08\% on YH,
87.54\% on PT and 75.52\% on RD). This aligns with our expectations for mixed tensor/node parallelism. 

Table~\ref{tbl:cache_miss_hierarch_node} shows cache miss and hit events on low degree graphs before and after applying hierarchical node parallelism (i.e., for pipelined node-level parallelism and hierarchical node parallelism). We see that both of the cache miss and hit events decrease, implying a better hit rate on the set of L2 caches (i.e., there aren't as many accesses to the L3 cache). Though the effects are not as significant as in the mixed parallelism case, as we illustrated, the hierarchical node parallelism 
is a conservative parallel strategy, which aims to reduce main memory access. Here $\text{Event}=\text{Access} = (\text{Miss} + \text{Hit})$, and we observe a significant ratio decrease in total accesses to the last level cache, where $R_{\text{Access}} = 82.81\%$ on MT, 83.17\% on MH, 73.66\% on AX,
68.13\% on YP, and 66.10\% on KA. Meanwhile, the ratio of LLC misses decreases by $74.28\%$ on MT, $76.36\%$ on MH, $65.68\%$ on AX,
$62.46\%$ on YP, and $66.76\%$ on KA. The above corroborates the notion that cache behavior is one of the mechanisms that benefit FrontGI's parallelism approaches.

\begin{table}[htbp]
\centering
\caption{LLC Hit/Miss counts For Low Degree Graphs.}
\label{tbl:cache_miss_hierarch_node}
\begin{tabular}{@{}cccccc@{}}
\toprule
& \multicolumn{2}{c}{\textbf{L3 hit}} & \multicolumn{2}{c}{\textbf{L3 miss}} \\ \cmidrule(l){2-5} 
& \textbf{Node} & \textbf{FrontGI-H} & \textbf{Node} & \textbf{FrontGI-H} \\ \midrule
MT & 1\,017\,M & 97\,M & 914\,M & 235\,M \\
MH & 5\,209\,M & 427\,M & 6\,608\,M & 1\,562\,M \\
AX & 712\,M & 93\,M & 1\,186\,M & 407\,M \\
YP & 2\,376\,M & 432\,M & 5\,731\,M & 2\,151\,M \\
KA & 463\,M & 166\,M & 1,360\,M & 452\,M \\
 \bottomrule
\end{tabular}
\end{table}

\section{Related Work}

Optimizing GNN training and inference performance across various hardware platforms has emerged as a popular research topic. Most of the previous work has been dedicated to enhancing GNN performance on GPUs and FPGAs (with some very recent work accelerating training on NPUs~\cite{ai2025neutronascend}), exploring techniques such as kernel fusion, memory access optimization, and customized accelerators. These efforts aim to address the challenges posed by the irregular memory access patterns and dataflow inherent to GNN workloads, making them better suited for high-performance inference and training on specialized hardware.

\subsection{GNNs on GPUs}

Sampling based GNN inference is the most widely used method to accelerate GNN inference on GPUs. In real world graphs, different vertices in a graph may share many common neighbors, causing redundant vertex feature vector access, loading and computation. 
GraphTensor~\cite{graphtensor_23} presents an efficient GNN processing framework by pipelining graph sampling, graph processing, and neural network execution behaviors according to the system's
hyperparameters, significantly shortening GNN execution latency.

To better leverage tensor core units in GNN computing on GPUs, computing systems need to remove redundant sparse representations and balance the load of tensor cores. TC-GNN~\cite{tcgnn_23} was proposed to translate sparse matrices to a denser representation, compressing large sparse matrices to small sparse blocks, thus finding a balance between sparse operations and high performance dense operations fixed by hardware. PckGNN~\cite{pckgnn_24} considers load balancing and data locality, categorizes graph matrices into different types with sparsity levels, enhancing load balance for sparser graphs and cache data reuse for denser graphs respectively, improving sparse matrices operation speed, thus improving the aggregation operator's efficiency. RT-GNN~\cite{rtgnn_24} exploits both graph sparsity and GPU computing cores (tensor cores and CUDA cores). It manages the intermediate aggregation results hierarchically to avoid redundant computation, partitioning graphs to graph blocks, then assigning graph blocks to different kinds of cores according to the blocks' sparsity, processing blocks with tensor cores and CUDA cores simultaneously. With the growth of graph size and diversity of GNN architecture, existing libraries are often not able to handle different workflow types of different GNNs. GNNAdvisor~\cite{gnna_21} was proposed to accelerate GNN inference by optimizing hierarchical GPU memory use utilizing the properties of input graphs and GNN architectures.

\subsection{GNNs on FPGAs}

FPGAs are a compelling solution for GNN inference acceleration due to their massive parallelism, better performance per watt than general-purpose computing platforms, reconfigurability for enhanced flexibility, and the ability to leverage concurrent data flow for efficient execution.
AWB-GCN~\cite{geng2020awb}, one of the earliest FPGA-based GCN accelerators, addresses workload imbalance caused by the distinct sparsity levels of different graph components using HDL. 
GenGNN~\cite{abi2022gengnn} introduces a framework for accelerating GNNs with real-time requirements by storing the entire graph and corresponding intermediate results in the on-chip resources of the FPGA. GNNHLS~\cite{zhao2023gnnhls} presents a comprehensive framework comprising both software and hardware components to accelerate various GNN models on FPGAs using High-Level Synthesis (HLS).
HLPerf~\cite{zhao2024hlperf} proposes a simulation-based performance evaluation framework for GNN inference, enabling early design space exploration and significantly reducing performance evaluation cycles.
The develop time for GNN acceleration on FPGAs remains non-trivial due to the excessive design space to be explored, prolonged compilation times, and the inherent differences from conventional software programming languages. In fact, the performance gap between a well-optimized and a naive implementation of the same kernel can span two to three orders of magnitude~\cite{sohrabizadeh2022autodse,fhx+22}.

\section{Conclusions and Future Work}
\label{sec:conclude}

Libraries such as DGL and PyG already offer an excellent high-productivity development environment for graph neural network applications. However, as the available execution platforms modernize, it is essential to explore how they implement the underlying computations.

As processor core counts scale up, the techniques needed to achieve high performance on many applications vary. This is clearly the case for inference on GNNs executing on high core-count modern processors. Motivated by the memory hierarchy of these systems, we have proposed a pair of parallelization techniques for GNN inference and have assessed their efficacy empirically using several GNN models and twelve representative graphs. The techniques we propose are straightforward to incorporate into any of the high-productivity environments designed for GNN inference.

Our experimental results show performance gains of up to $3.35\times$ relative to the higher performing of two commonly used standard libraries. In addition, which parallelization technique is better depends in large part on the topological properties of the graph.  In our case, graphs with low average degree ($<$ 20) scaled better using the hierarchical node parallelism technique, graphs with high average degree ($>$ 50) scaled better using the mixed tensor/node parallelism technique and for graphs whose average degree falls in the range of 20 to 50, either parallelization strategy of FrontGI proves effective.

This work can be extended in a number of ways. First, we explored scaling up the inference of GCN, GraphSAGE, GIN, and SGC. However, additional models worth investigation are described in a recent review by Khemani et al.~\cite{khemani2024review}, e.g., parallelizing graph attention in GAT~\cite{velivckovic2017gat} with our method. Second, while this work is focused on inference, there is a possibly larger need for acceleration of GNN training, where automatically selecting efficient parallel strategies may provide increasing benefits as the number of cores in a system grows. Before one incorporates FrontGI approaches into production libraries such as DGL or PyG, its effectiveness across a broader set of circumstances (both GNN models and graph datasets) is required.


\bibliography{refs}% common bib file

@article{rtgnn_24,
  title={{RT-GNN}: Accelerating Sparse Graph Neural Networks by {Tensor-CUDA} Kernel Fusion},
  author={Yan, Jianrong and Jiang, Wenbin and He, Dongao and Wen, Suyang and Li, Yang and Jin, Hai and Shao, Zhiyuan},
  journal={ACM Trans.\ on Architecture and Code Optimization},
  year={2025},
  month = {March},
  volume = 22,
  number = 1,
  doi={10.1145/3702001},
  pagesx = {32:1-32:27},
  pages = {32},
  organization={ACM},
  publisher={ACM New York, NY}
}

@article{goto2008anatomy,
  title={Anatomy of high-performance matrix multiplication},
  author={Goto, Kazushige and {van de Geijn}, Robert A},
  journal={ACM Transactions on Mathematical Software (TOMS)},
  volume={34},
  number={3},
  pages={1--25},
  year={2008},
  doi={10.1145/1356052.1356053},
  organization={ACM},
  publisher={ACM New York, NY, USA}
}

@inproceedings{tcgnn_23,
  title={{TC-GNN}: Bridging Sparse {GNN} Computation and Dense Tensor Cores on {GPUs}},
  author={Wang, Yuke and Feng, Boyuan and Wang, Zheng and Huang, Guyue and Ding, Yufei},
  booktitle={USENIX Annual Technical Conference},
  pages={149--164},
  year={2023}
}

@inproceedings{pckgnn_24,
  title={{PckGNN}: Optimizing Aggregation Operators with Packing Strategies in Graph Neural Networks},
  author={Hu, Zhengding and Sun, Jingwei and Li, Zhongyang and Sun, Guangzhong},
  booktitle={Int'l Parallel and Distributed Processing Symp.},
  pages={2--13},
  year={2024},
  doi={10.1109/IPDPS57955.2024.00010},
  organization={IEEE}
}

@inproceedings{graphtensor_23,
  title={{GraphTensor}: comprehensive {GNN}-acceleration framework for efficient parallel processing of massive datasets},
  author={Jang, Junhyeok and Kwon, Miryeong and Gouk, Donghyun and Bae, Hanyeoreum and Jung, Myoungsoo},
  booktitle={Int'l Parallel and Distributed Processing Symp.},
  pages={2--12},
  year={2023},
  doi={10.1109/IPDPS54959.2023.00011},
  organization={IEEE}
}

@inproceedings{gnna_21,
  title={{GNNAdvisor}: An adaptive and efficient runtime system for {GNN} acceleration on {GPUs}},
  author={Wang, Yuke and Feng, Boyuan and Li, Gushu and Li, Shuangchen and Deng, Lei and Xie, Yuan and Ding, Yufei},
  booktitle={15th USENIX Symposium on Operating Systems Design and Implementation},
  pages={515--531},
  year={2021}
}

@misc{gcn,
  title={Semi-supervised classification with graph convolutional networks},
  author={Kipf, Thomas N and Welling, Max},
  howpublished={arXiv preprint arXiv:1609.02907},
  doi={10.48550/arXiv.1609.02907},
  year={2016}
}

@inproceedings{bfs23,
  title={Traversing Large Compressed Graphs on {GPUs}},
  author={Gera, Prasun and Kim, Hyesoon},
  booktitle={Int'l Parallel and Distributed Processing Symposium},
  pages={25--35},
  year={2023},
  doi={10.1109/IPDPS54959.2023.00013},
  organization={IEEE}
}

@inproceedings{cusha14,
  title={{CuSha}: Vertex-centric graph processing on {GPUs}},
  author={Khorasani, Farzad and Vora, Keval and Gupta, Rajiv and Bhuyan, Laxmi N},
  booktitle={23rd Int'l Symp.\ on High-performance Parallel and Dist.\ Computing},
  pages={239--252},
  year={2014},
  doi={10.1145/2600212.2600227},
  organization={ACM}
}

@inproceedings{disk12,
  title={{GraphChi}: {Large-Scale} graph computation on just a {PC}},
  author={Kyrola, Aapo and Blelloch, Guy and Guestrin, Carlos},
  booktitle={10th USENIX Symposium on Operating Systems Design and Implementation},
  pages={31--46},
  year={2012}
}

@misc{megatron,
  title={{Megatron-LM}: Training multi-billion parameter language models using model parallelism},
  author={Shoeybi, Mohammad and Patwary, Mostofa and Puri, Raul and LeGresley, Patrick and Casper, Jared and Catanzaro, Bryan},
  howpublished={arXiv preprint arXiv:1909.08053},
  doi={10.48550/arXiv.1909.08053},
  year={2019}
}

@inproceedings{tesseract22,
  title={Tesseract: Parallelize the tensor parallelism efficiently},
  author={Wang, Boxiang and Xu, Qifan and Bian, Zhengda and You, Yang},
  booktitle={51st Int'l Conference on Parallel Processing},
  pagesx={12:1--12:11},
  doi={10.1145/3545008.3545087},
  year={2022},
  organization={ACM}
}

@article{pytorch,
  title={{PyTorch}: An imperative style, high-performance deep learning library},
  authorx={Paszke, Adam and Gross, Sam and Massa, Francisco and Lerer, Adam and Bradbury, James and Chanan, Gregory and Killeen, Trevor and Lin, Zeming and Gimelshein, Natalia and Antiga, Luca and others},
  author = {Paszke, Adam and others},
  journal={Advances in Neural Information Processing Systems},
  volume={32},
  year={2019}
}

@inproceedings{distgnn21,
  title={{DistGNN}: Scalable distributed training for large-scale graph neural networks},
  author={{Vasimuddin Md} and Misra, Sanchit and Ma, Guixiang and Mohanty, Ramanarayan and Georganas, Evangelos and Heinecke, Alexander and Kalamkar, Dhiraj and Ahmed, Nesreen K and Avancha, Sasikanth},
  booktitle={Int'l Conference for High Performance Computing, Networking, Storage and Analysis},
  pagesx={76:1--76:14},
  year={2021},
  doi={10.1145/3458817.3480856},
  organization={ACM}
}

@inproceedings{lin2024argo,
  title={{ARGO}: An auto-tuning runtime system for scalable {GNN} training on multi-core processor},
  author={Lin, Yi-Chien and Chen, Yuyang and Gobriel, Sameh and Jain, Nilesh and Jha, Gopi Krishna and Prasanna, Viktor},
  booktitle={International Parallel and Distributed Processing Symposium},
  pages={361--372},
  year={2024},
  doi={10.1109/IPDPS57955.2024.00039},
  organization={IEEE}
}

@inproceedings{gong2022graphite,
  title={Graphite: optimizing graph neural networks on {CPUs} through cooperative software-hardware techniques},
  author={Gong, Zhangxiaowen and Ji, Houxiang and Yao, Yao and Fletcher, Christopher W and Hughes, Christopher J and Torrellas, Josep},
  booktitle={49th Int'l Symposium on Computer Architecture},
  pages={916--931},
  doi={10.1145/3470496.3527403},
  year={2022},
  organization={ACM}
}

@inproceedings{fu2021automatic,
  title={Automatic generation of high-performance inference kernels for graph neural networks on multi-core systems},
  author={Fu, Qiang and Huang, H Howie},
  booktitle={50th Int'l Conf.\ on Parallel Processing},
  pagesx={33:1--33:11},
  year={2021},
  doi={10.1145/3472456.3473511},
  organization={ACM}
}

@inproceedings{zhao2023gnnhls,
  title={{GNNHLS}: Evaluating Graph Neural Network Inference via High-Level Synthesis},
  author={Zhao, Chenfeng and Dong, Zehao and Chen, Yixin and Zhang, Xuan and Chamberlain, Roger D},
  booktitle={41st Int'l Conference on Computer Design},
  pages={574--577},
  year={2023},
  month = {November},
  doi={10.1109/ICCD58817.2023.00092},
  organization={IEEE}
}

@article{han2022vision,
  title={Vision {GNN}: An image is worth graph of nodes},
  author={Han, Kai and Wang, Yunhe and Guo, Jianyuan and Tang, Yehui and Wu, Enhua},
  journal={Advances in Neural Information Processing Systems},
  volume={35},
  pages={8291--8303},
  year={2022}
}

@inproceedings{wu2021deep,
  title={Deep learning on graphs for natural language processing},
  author={Wu, Lingfei and Chen, Yu and Ji, Heng and Liu, Bang},
  booktitle={Int'l Conf.\ on Research and Development in Information Retrieval},
  pages={2651--2653},
  year={2021},
  doi={10.1145/3404835.3462809},
  organization = {ACM}
}

@misc{pyg19,
  title={Fast graph representation learning with {PyTorch} {Geometric}},
  author={Fey, Matthias and Lenssen, Jan Eric},
  howpublished={arXiv preprint arXiv:1903.02428},
  doi={10.48550/arXiv.1903.02428},
  year={2019}
}

@inproceedings{dgl19,
  title={{Deep Graph Library}: Towards efficient and scalable deep learning on graphs},
  author={Wang, Minjie Yu},
  booktitle={ICLR Workshop on Representation Learning on Graphs and Manifolds},
  year={2019}
}

@article{wu2024can,
  title={Can Graph Learning Improve Planning in {LLM}-based Agents?},
  author={Wu, Xixi and Shen, Yifei and Shan, Caihua and Song, Kaitao and Wang, Siwei and Zhang, Bohang and Feng, Jiarui and Cheng, Hong and Chen, Wei and Xiong, Yun and others},
  journal={Advances in Neural Information Processing Systems},
  volume={37},
  pages={5338--5383},
  year={2024}
}

@inproceedings{geng2020awb,
  title={{AWB-GCN}: A graph convolutional network accelerator with runtime workload rebalancing},
  author={Geng, Tong and Li, Ang and Shi, Runbin and Wu, Chunshu and Wang, Tianqi and Li, Yanfei and Haghi, Pouya and Tumeo, Antonino and Che, Shuai and Reinhardt, Steve and Martin C. Herbordt},
  booktitle={53rd Int'l Symp.\ on Microarchitecture},
  pages={922--936},
  year={2020},
  doi={10.1109/MICRO50266.2020.00079},
  organization={IEEE}
}

@article{zhao2024hlperf,
  title={{HLPerf}: Demystifying the Performance of {HLS}-based Graph Neural Networks with Dataflow Architectures},
  author={Zhao, Chenfeng and Faber, Clayton J and Chamberlain, Roger D and Zhang, Xuan},
  journal={ACM Transactions on Reconfigurable Technology and Systems},
  volume={18},
  number={1},
  pagesx={2:1--2:26},
  pages={2},
  year={2025},
  month={March},
  doi={10.1145/3655627},
  organization={ACM},
  publisher={ACM New York, NY}
}

@article{sohrabizadeh2022autodse,
  title={{AutoDSE}: Enabling Software Programmers to Design Efficient {FPGA} Accelerators},
  author={Sohrabizadeh, Atefeh and Yu, Cody Hao and Gao, Min and Cong, Jason},
  journal={ACM Transactions on Design Automation of Electronic Systems},
  volume={27},
  number={4},
  pagesx={32:1--32:27},
  pages={32},
  year={2022},
  organization={ACM},
  doi={10.1145/3494534}
}

@inproceedings{fhx+22,
  author = {Clayton J. Faber and Steven D. Harris and Zhili Xiao and Roger D. Chamberlain and Anthony M. Cabrera},
  title = {Challenges Designing for {FPGAs} Using High-Level Synthesis},
  booktitle = {Proc.\ of High-Performance Extreme Computing Conference},
  month = {September},
  year = 2022,
  numpages = 7,
  doi = {10.1109/HPEC55821.2022.9926398},
  organization = {IEEE}
}

@misc{abi2022gengnn,
  title={{GenGNN}: A Generic {FPGA} Framework for Graph Neural Network Acceleration},
  author={Abi-Karam, Stefan and He, Yuqi and Sarkar, Rishov and Sathidevi, Lakshmi and Qiao, Zihang and Hao, Cong},
  howpublished={arXiv preprint arXiv:2201.08475},
  doi={10.48550/arXiv.2201.08475},
  year={2022}
}

@inproceedings{zhao2023supercut,
  title={{SuperCut}: Communication-Aware Partitioning for Near-Memory Graph Processing},
  author={Zhao, Chenfeng and Chamberlain, Roger D and Zhang, Xuan},
  booktitle={20th Int'l Conference on Computing Frontiers},
  pages={42--51},
  organization = {ACM},
  year={2023},
  month={May},
  doi={10.1145/3587135.3592209}
}

@article{clauset2009power,
  title={Power-law distributions in empirical data},
  author={Clauset, Aaron and Shalizi, Cosma Rohilla and Newman, Mark EJ},
  journal={SIAM Review},
  volume={51},
  number={4},
  pages={661--703},
  year={2009},
  doi={10.1137/070710111},
  publisher={SIAM}
}

@article{gupta1993performance,
  title={Performance properties of large scale parallel systems},
  author={Gupta, Anshul and Kumar, Vipin},
  journal={Journal of Parallel and Distributed Computing},
  volume={19},
  number={3},
  pages={234--244},
  year={1993},
  doi={10.1006/jpdc.1993.1107},
  organization={Elsevier}
}

@article{hu2020open,
  title={{Open Graph Benchmark}: Datasets for machine learning on graphs},
  author={Hu, Weihua and Fey, Matthias and Zitnik, Marinka and Dong, Yuxiao and Ren, Hongyu and Liu, Bowen and Catasta, Michele and Leskovec, Jure},
  journal={Advances in Neural Information Processing Systems},
  volume={33},
  pages={22118--22133},
  year={2020}
}

@inproceedings{ma2019neugraph,
  title={{NeuGraph}: Parallel deep neural network computation on large graphs},
  author={Ma, Lingxiao and Yang, Zhi and Miao, Youshan and Xue, Jilong and Wu, Ming and Zhou, Lidong and Dai, Yafei},
  booktitle={USENIX Annual Technical Conference},
  pages={443--458},
  year={2019}
}

@article{dongarra1990set,
  title={A set of level 3 basic linear algebra subprograms},
  author={Dongarra, Jack J and Du Croz, Jeremy and Hammarling, Sven and Duff, Iain S},
  journal={ACM Transactions on Mathematical Software},
  volume={16},
  number={1},
  pages={1--17},
  year={1990},
  doi={10.1145/77626.79170},
  organization={ACM},
  publisher={ACM New York, NY, USA}
}

@article{khemani2024review,
  title={A review of graph neural networks: concepts, architectures, techniques, challenges, datasets, applications, and future directions},
  author={Khemani, Bharti and Patil, Shruti and Kotecha, Ketan and Tanwar, Sudeep},
  journal={Journal of Big Data},
  volume={11},
  number={1},
  pages={18:1-18:42},
  year={2024},
  doi={10.1186/s40537-023-00876-4},
  organization={Springer}
}

@misc{zeng2019yelpgraph,
  title={{GraphSAINT}: Graph sampling based inductive learning method},
  author={Zeng, Hanqing and Zhou, Hongkuan and Srivastava, Ajitesh and Kannan, Rajgopal and Prasanna, Viktor},
  howpublished={arXiv preprint arXiv:1907.04931},
  doi={10.48550/arXiv.1907.04931},
  year={2019}
}

@inproceedings{murphy2010graph500,
  title={Introducing the {Graph 500}},
  author={Murphy, Richard C and Wheeler, Kyle B and Barrett, Brian W and Ang, James A},
  booktitle={Cray Users Group},
  volumex={19},
  numberx={45-74},
  pagesx={22},
  month={May},
  year={2010}
}

@inproceedings{kunegis2013konect,
  title={{KONECT}: the {Koblenz} network collection},
  author={Kunegis, J{\'e}r{\^o}me},
  booktitle={22nd Int'l Conference on World Wide Web},
  pages={1343--1350},
  year={2013},
  doi={10.1145/2487788.2488173},
  organization={ACM}
}

@inproceedings{dror2012yahoo,
  title={{The Yahoo! Music Dataset and KDD-Cup’11}},
  author={Dror, Gideon and Koenigstein, Noam and Koren, Yehuda and Weimer, Markus},
  booktitle={Proceedings of KDD Cup 2011},
  pages={3--18},
  year={2012},
  organization={PMLR}
}

@article{lewis2004rtdata,
  title={{RCV1}: A new benchmark collection for text categorization research},
  author={Lewis, David D and Yang, Yiming and Rose, Tony G and Li, Fan},
  journal={Journal of Machine Learning Research},
  volume={5},
  numberx={Apr},
  pages={361--397},
  year={2004}
}

@article{hamilton2017sage,
  title={Inductive representation learning on large graphs},
  author={Hamilton, Will and Ying, Zhitao and Leskovec, Jure},
  journal={Advances in Neural Information Processing Systems},
  volume={30},
  year={2017}
}

@misc{xu2018gin,
  title={How powerful are graph neural networks?},
  author={Xu, Keyulu and Hu, Weihua and Leskovec, Jure and Jegelka, Stefanie},
  howpublished={arXiv preprint arXiv:1810.00826},
  doi={10.48550/arXiv.1810.00826},
  year={2018}
}

@inproceedings{nassif2022sapphire,
  title={{Sapphire Rapids}: The next-generation {Intel Xeon} scalable processor},
  author={Nassif, Nevine and Munch, Ashley O and Molnar, Carleton L and Pasdast, Gerald and Lyer, Sitaraman V and Yang, Zibing and Mendoza, Oscar and Huddart, Mark and Venkataraman, Srikrishnan and Kandula, Sireesha and others},
  booktitle={Int'l Solid-State Circuits Conference},
  volume={65},
  pages={44--46},
  year={2022},
  doi={10.1109/ISSCC42614.2022.9731107},
  organization={IEEE}
}

@article{bhargava2024amd,
  title={{AMD Next-Generation ``Zen 4'' Core and 4th Gen AMD EPYC Server CPUs}},
  author={Bhargava, Ravi and Troester, Kai},
  journal={IEEE Micro},
  volume={44},
  number={3},
  pages={8--17},
  year={2024},
  doi={10.1109/MM.2024.3375070},
  organization={IEEE}
}

@inproceedings{smith202411amdgpu,
  title={{AMD Instinct\textsuperscript{TM} MI300 series modular chiplet package--HPC and AI accelerator for exa-class systems}},
  author={Smith, Alan and Chapman, Eric and Patel, Chintan and Swaminathan, Raja and Wuu, John and Huang, Tyrone and Jung, Wonjun and Kaganov, Alexander and McIntyre, Hugh and Mangaser, Ramon},
  booktitle={Int'l Solid-State Circuits Conference},
  volume={67},
  pages={490--492},
  year={2024},
  doi={10.1109/ISSCC49657.2024.10454441},
  organization={IEEE}
}

@inproceedings{jzc25,
  author = {Binglin Ji and Chenfeng Zhao and Roger D. Chamberlain},
  title = {{FGI}: Fast {GNN} Inference on Multi-Core Systems},
  booktitle = {IEEE Int'l Parallel and Distributed Processing Symposium Workshops},
  month = {June},
  year = 2025,
  eprintx = {jzc25.pdf},
  doi = {10.1109/IPDPSW66978.2025.00119},
  pages = {748-757},
  organization={IEEE},
  notex = {Presented at Workshop on Graphs, Architectures, Programming, and Learning, Milan, Italy.}
}

@article{li2020chiplet,
  title={Chiplet heterogeneous integration technology—Status and challenges},
  author={Li, Tao and Hou, Jie and Yan, Jinli and Liu, Rulin and Yang, Hui and Sun, Zhigang},
  journal={Electronics},
  volume={9},
  number={4},
  pages={670},
  year={2020},
  doi={10.3390/electronics9040670},
  organization={MDPI}
}

@article{weisfeiler1968reduction,
  title={The reduction of a graph to canonical form and the algebra which appears therein},
  author={Weisfeiler, Boris and Leman, Andrei},
  journal={Nauchno-Technicheskaya Informatsia},
  volume={2},
  number={9},
  pages={12--16},
  year={1968}
}

@inproceedings{wu2019sgc,
  title={Simplifying graph convolutional networks},
  author={Wu, Felix and Souza, Amauri and Zhang, Tianyi and Fifty, Christopher and Yu, Tao and Weinberger, Kilian},
  booktitle={International Conference on Machine Learning},
  pages={6861--6871},
  year={2019},
  organization={PMLR}
}

@article{velivckovic2017gat,
  title={Graph attention networks},
  author={Veli{\v{c}}kovi{\'c}, Petar and Cucurull, Guillem and Casanova, Arantxa and Romero, Adriana and Lio, Pietro and Bengio, Yoshua},
  journal={arXiv preprint arXiv:1710.10903},
  doi={10.48550/arXiv.1710.10903},
  year={2017}
}

@inproceedings{dong2023cktgnn,
title={Ckt{GNN}:  Circuit Graph Neural Network for Electronic Design Automation},
author={Zehao Dong and Weidong Cao and Muhan Zhang and Dacheng Tao and Yixin Chen and Xuan Zhang},
booktitle={Int'l Conference on Learning Representations},
year={2023},
urlx={https://openreview.net/forum?id=NE2911Kq1sp}
}

@ARTICLE{fan2022graph,
  author={Fan, Wenqi and Ma, Yao and Li, Qing and Wang, Jianping and Cai, Guoyong and Tang, Jiliang and Yin, Dawei},
  journal={IEEE Transactions on Knowledge and Data Engineering}, 
  title={A Graph Neural Network Framework for Social Recommendations}, 
  year={2022},
  volume={34},
  number={5},
  pages={2033-2047},
  organization={IEEE},
  doi={10.1109/TKDE.2020.3008732}
}

@article{wu2022graph,
author = {Wu, Shiwen and Sun, Fei and Zhang, Wentao and Xie, Xu and Cui, Bin},
title = {Graph Neural Networks in Recommender Systems: A Survey},
year = {2022},
organization = {ACM},
address = {New York, NY, USA},
volume = {55},
number = {5},
doi = {10.1145/3535101},
journal = {ACM Comput. Surv.},
month = {December},
pagesx = {97:1-97:37},
pages = {97}
}

@article{zhang2021graph,
  title={Graph neural networks and their current applications in bioinformatics},
  author={Zhang, Xiao-Meng and Liang, Li and Liu, Lin and Tang, Ming-Jing},
  journal={Frontiers in Genetics},
  volume={12},
  pages={690049},
  year={2021},
  organization={Frontiers Media SA},
  doi={10.3389/fgene.2021.690049}
}

@article{tabassum2018social,
  title={Social network analysis: An overview},
  author={Tabassum, Shazia and Pereira, Fabiola SF and Fernandes, Sofia and Gama, Jo{\~a}o},
  journal={Wiley Interdisciplinary Reviews: Data Mining and Knowledge Discovery},
  volume={8},
  number={5},
  pages={e1256},
  year={2018},
  organization={Wiley},
  doi={10.1002/widm.1256}
}

@article{whaley2001automated,
  title={Automated empirical optimizations of software and the {ATLAS} project},
  author={Whaley, R Clint and Petitet, Antoine and Dongarra, Jack J},
  journal={Parallel computing},
  volume={27},
  number={1-2},
  pages={3--35},
  year={2001},
  doi={10.1016/S0167-8191(00)00087-9},
  organization={Elsevier}
}

@article{ai2025neutronascend,
  title={{NeutronAscend}: Optimizing {GNN} Training with {Ascend AI} Processors},
  author={Ai, Xin and Zhang, Bing and Wang, Qiange and Zhang, Yanfeng and Yuan, Hao and Gong, Shufeng and Yu, Ge},
  journal={ACM Transactions on Architecture and Code Optimization},
  volume={22},
  number={4},
  pagesx={123:1--123:26},
  pages={123},
  year={2025},
  organization={ACM},
  publisher={ACM New York, NY},
  doi={10.1145/3762662}
}

@article{chen2026quantitative,
  title={Quantitative Analysis and Performance Optimization of Graph Neural Networks on Multi-core {CPUs}},
  author={Chen, Kangkang and Su, Huayou and Yang, Xi and An, Zitong and Dou, Yong and Li, Dongsheng},
  journal={ACM Transactions on Architecture and Code Optimization},
  volume={23},
  number={2},
  pagesx={53:1-53:25},
  pages={53},
  year={2026},
  organization={ACM},
  publisher={ACM New York, NY},
  doi={10.1145/3807957}
}

@article{ozsoy2020comprehensive,
  title={A comprehensive performance comparison of dedicated and embedded {GPU} systems},
  author={{\"O}zsoy, Adnan},
  journal={Dicle {\"U}niversitesi M{\"u}hendislik Fak{\"u}ltesi M{\"u}hendislik Dergisi},
  volume={11},
  number={3},
  pages={1011--1020},
  year={2020},
  organization={Dicle University},
  doi={10.24012/dumf.742732}
}

\end{document}